\documentclass[useAMS,usenatbib]{mn2e}
\usepackage{url}
\usepackage{stmaryrd}
\usepackage[pdftex]{graphicx}
\usepackage[draft]{hyperref}
\usepackage[flushleft]{threeparttable}

\def\aj{AJ}%
\def\apj{ApJ}%
\def\apjl{ApJ}%
\def\apjs{ApJS}%
\def\aap{A\&A}%
\def\mnras{MNRAS}%
\def\pasp{PASP}%
\def\nat{Nature}%
\def\procspie{Proc.~SPIE}%

\def\hi{$H{\sc I}$}

\begin{document}

\title[VIMOS Observations of an Occulting Galaxy Pair]{VLT/VIMOS Observations of an Occulting Galaxy Pair:\\ Redshifts and Effective Extinction Curve}

\author[Holwerda et al.]{B.W. Holwerda$^{1}$\thanks{E-mail:
benne.holwerda@esa.int}, 
T. B\"oker$^{1}$,
J.J. Dalcanton$^{2}$, 
W.C. Keel$^{3}$
and R.S. de Jong$^{4}$\\
$^{1}$ European Space Agency Research Fellow (ESTEC), Keplerlaan 1, 2200 AG Noordwijk, The Netherlands \\
$^{2}$ Department of Astronomy, University of Washington, Box 351580, Seattle, WA 98195, USA\\
$^{3}$ Department of Physics and Astronomy, University of Alabama, Box 870324, Tuscaloosa, AL 35487, USA\\
$^{4}$ Leibniz Institut f\"ur Astrophysik Potsdam (AIP), An der Sternwarte 16, 14482 Potsdam, Germany}

\date{Accepted {\em pending}. Received 2011 September; in original form 2011 September }

\pagerange{\pageref{firstpage}--\pageref{lastpage}} \pubyear{2010}

\maketitle

\label{firstpage}

\begin{abstract}
We present VLT/VIMOS IFU observations of an occulting galaxy pair previously discovered in HST observations. The foreground galaxy is a low-inclination spiral disk, which causes clear attenuation features seen agains the bright bulge and disk of the background galaxy. We find redshifts of $z=0.064 \pm0.003$ and z=0.065 for the foreground and background galaxy respectively. This relatively small difference does not rule out gravitational interaction between the two galaxies. Emission line ratios point to a star-forming, not AGN-dominated foreground galaxy.

We fit the Cardelli, Clayton \& Mathis (CCM) extinction law to the spectra of individual fibres to derive slope  ($R_V$) and normalization ($A_V$). The normalization agrees with the HST attenuation map and the slope is lower than the Milky Way relation ($R_V<3.1$), which is likely linked to the spatial sampling of the disk. We speculate that the values of $R_V$ point to either coherent ISM structures in the disk larger than usual ($\sim9$ kpc) or higher starting values of $R_V$, indicative of recent processing of the dust.

The foreground galaxy is a low stellar mass spiral ($M_* \sim 3 \times 10^9 M_\odot$) with a high dust content ($M_{\rm dust} \sim 0.5 \times 10^6 M_\odot$). 
The dust disk geometry visible in the HST image would explain the observed SED properties of smaller galaxies: a lower mean dust temperature, a high dust-to-stellar mass ratio but relatively little optical attenuation. Ongoing efforts to find occulting pairs with a small foreground galaxies will show how common this geometry is.
\end{abstract}


\section{\label{s:intro}Introduction}

The distribution and properties of interstellar dust in spiral galaxies is still the dominant unknown in models of the spectral energy distribution (SED) of spiral galaxies \citep[e.g.,][]{Popescu00, Popescu05rev, Pierini04, Baes03, Baes10a, Bianchi00c, Bianchi07, Bianchi08, Driver08, Jonsson10, Holwerda12a}. Recently, the focus has shifted to how the dust is distributed, and its impact on SED models and distance measurements \citep[e.g., SNIa,][]{Holwerda08a}. 

Knowledge of the dust geometry in external spiral galaxies has seen great improvements with the {\em Herschel Space Observatory}. This new facility has mapped the temperature gradient in the dusty ISM \citep{Bendo10b,Smith10b, Engelbracht10,Pohlen10, Foyle12}, notably in the spiral arms  \citep{Bendo10b} and bulge \citep{Engelbracht10} of nearby galaxies, and the power spectrum of the ISM, gas and dust \citep{Combes12}. However, the smallest scales can still only be probed in some of the most nearby galaxies, e.g., Andromeda, the Magellanic Clouds or the Milky Way itself. In the case of smaller galaxies, {\em Spitzer Space Telescope} and {\em Herschel Space Observatory} results are restricted to a few individual galaxies \citep[e.g.,][]{Hinz06, Hinz12, Hermelo13}, and those from stacked SEDs \citep{Popescu02b, Grossi10,Bourne12a}. 
Thus, observations in the optical or ultra-violet wavelength regimes can fill an observational gap to estimate the dust geometry in spiral disks or dwarf galaxies independently from sub-mm observations

\begin{figure*}
\begin{center}
\includegraphics[width=0.49\textwidth]{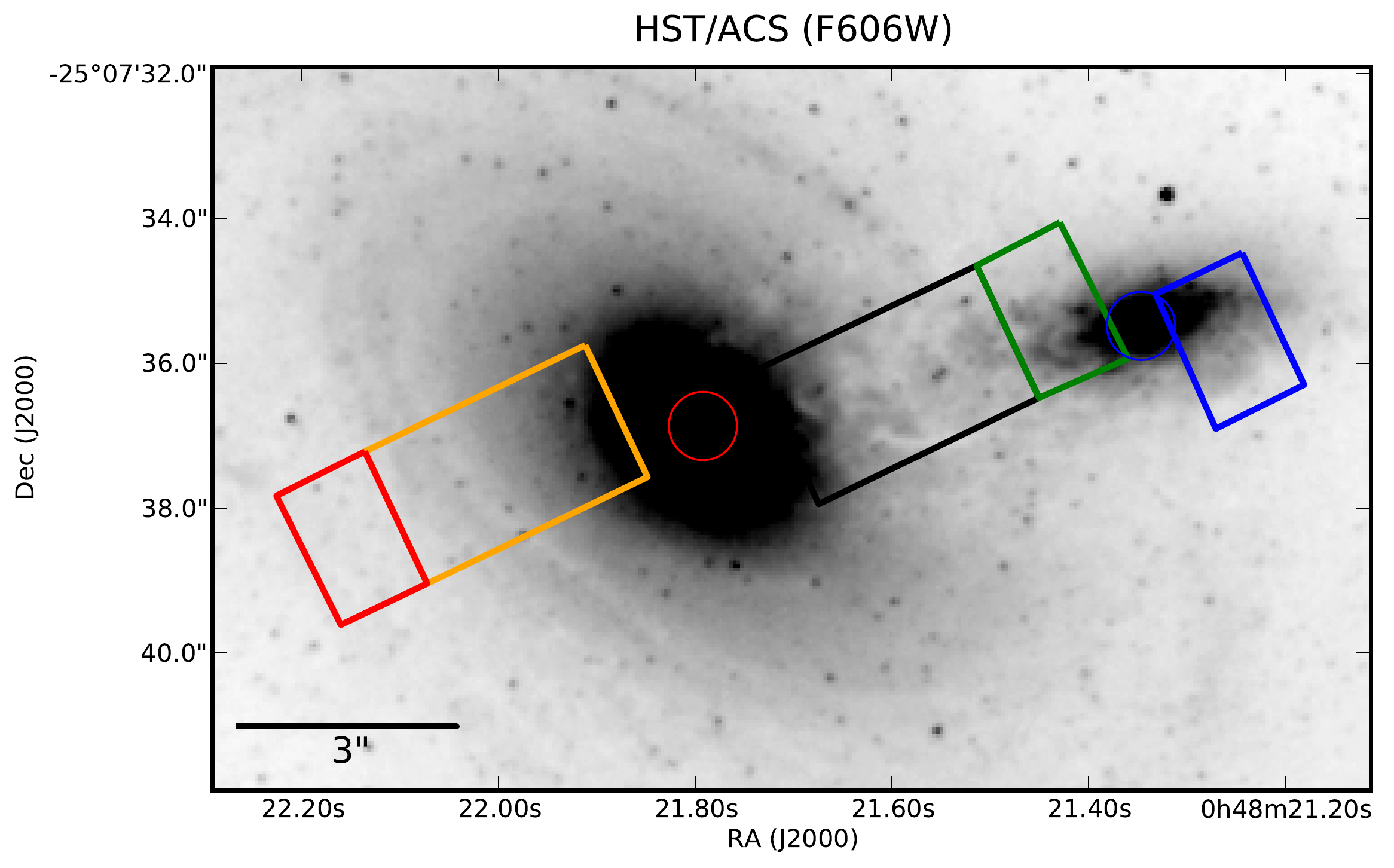}
\includegraphics[width=0.49\textwidth]{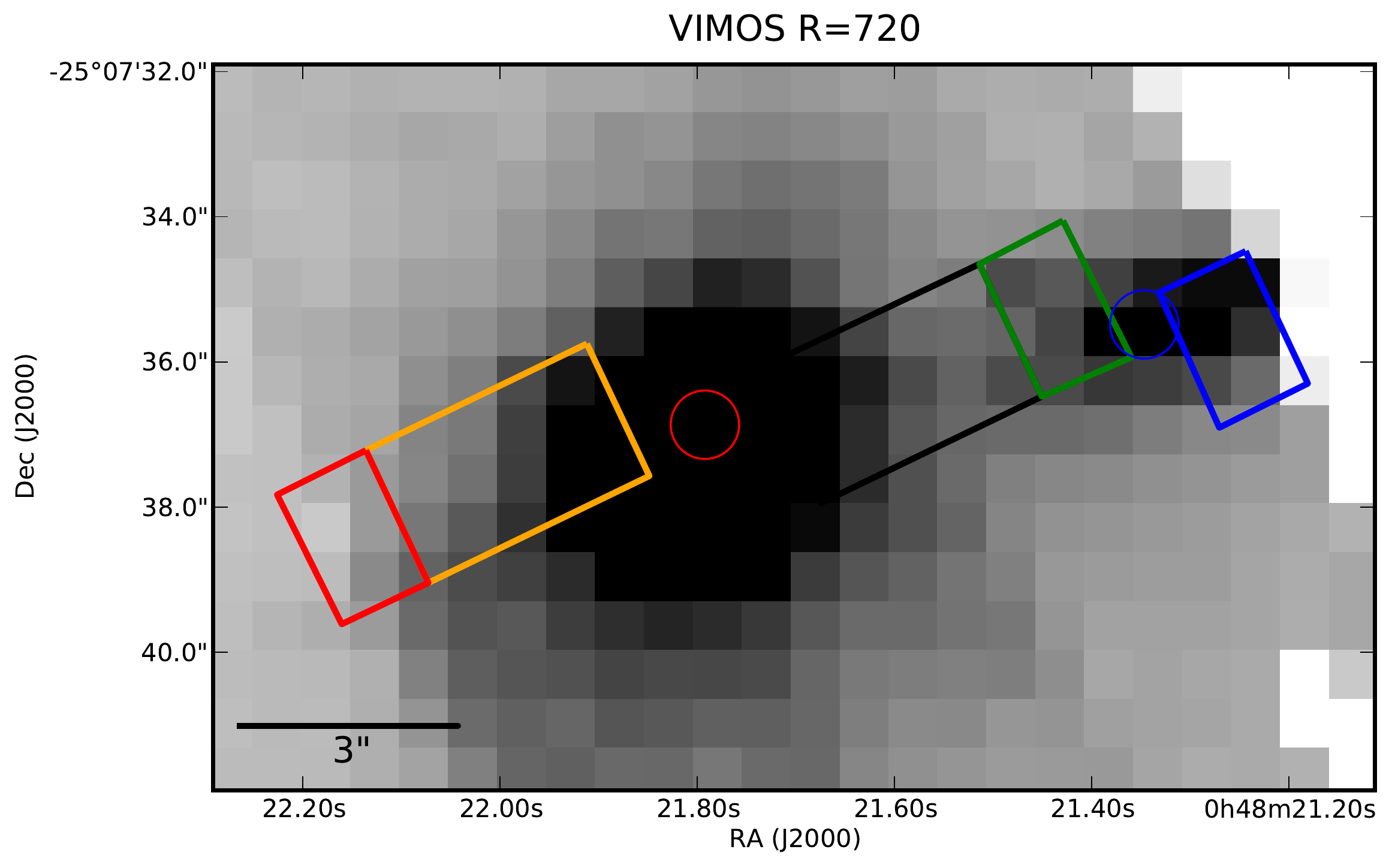}
\includegraphics[width=\textwidth]{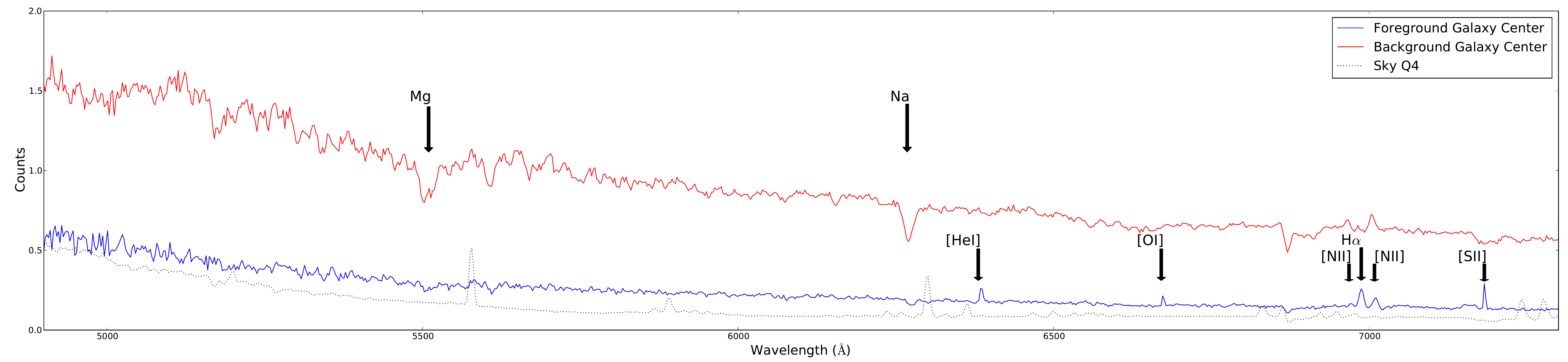}

\caption{The {\em HST/F606W} image (left) and the {\em VLT/VIMOS} reconstructed image (right) of the occulting galaxy pair (top two panels) and the sky-subtracted spectra of both nuclei (bottom panel). The blue and red spectrum correspond to the fibres on the centers of the foreground ($FG$, blue circle) and background galaxy ($BG$, red circle), respectively. The dotted spectrum is the sky contribution estimated from several ``empty" fibres in the fourth quadrant of the VIOS IFU. This sky contribution was subtracted from both FG and BG spectra. We analyze two apertures of the overlap region; one in which the foreground galaxy contributes a significant amount of flux (I, green box) and one for which we assume no foreground disk emission (II, black box). 
The flux contributions to Aperture I are estimated from the corresponding fibre positions for the foreground galaxy ($F'$, blue box) and background galaxy ($B'$, red box). Optical depths are calculated from equation \ref{eq:1}. Optical depth estimates for Aperture II (black box) are estimated from the ratio of flux in the aperture and the corresponding fibres on the other side of the background galaxy (orange box) and equation \ref{eq:2}.}
\label{f:fov}
\end{center}
\end{figure*}

To map dust extinction or attenuation\footnote{Because the actual contribution of scattered light is unknown, we will use the term dust attenuation throughout the paper, except where discussing the extinction-reddening relations, i.e., the observed extinction curves and law.}, one needs a known background source of light. In the case of a spiral galaxy overlapping a more distant galaxy, the more distant galaxy provides such a background light source. Estimating dust attenuation and mass from differential photometry in occulting pairs of galaxies was first proposed by \cite{kw92}. Their technique was then applied to all known pairs using ground-based optical images \citep{Andredakis92, Berlind97, kw99a, kw00a}, spectroscopy \citep{kw00b}, and later space-based {\em HST} images \citep[][]{kw01a, kw01b, Elmegreen01, Holwerda09}. These results agree with distant galaxy counts through spiral disks in {\em HST} images, an alternative approach \citep{Holwerda05,Holwerda05a,Holwerda05b,Holwerda05c,Holwerda05d,Holwerda05e,Holwerda12c}. 
More recently, additional pairs were found in the SDSS spectroscopic catalog \citep[86 pairs in][]{Holwerda07c}, and in an effort in the Galaxy Zoo project \citep{galaxyzoo}; the current count is 1993 pairs \citep{Keel13}. 
\cite{Holwerda09} (hereafter H09) have recently found a superb example of an overlapping pair in Hubble Space Telescope ({\em HST}) data (Figure \ref{f:fov}).

This system is one of the best examples of an occulting pair: it combines an optimal viewing configuration (face-on foreground disk and partial overlap) and a bright background galaxy. This pair was serendipitously observed with the Advanced Camera for Surveys ({\em ACS})  in observations of the outskirts and halo of NGC 253 \citep{Bailin11} as part of the ACS Nearby Galaxies Survey Treasury \citep[ANGST,][]{angst}. 
Our analysis in H09  showed an extended dusty disk in the foreground galaxy, extending well beyond its optical size. The average reddening-attenuation relation appears to be close to the Milky Way Extinction Law ($R_V = (A_B - A_V) / A_V = 3.1$), although with significant scatter (Figure 2 in H09). This measurement was based on the broad-band {\em HST} filters ({\em F475W, F606W} and {\em F814W}) and averaged over the entire overlap region. The attenuation is largely clustered in a series of spiral structures (Figure 2 in H09). Such an extended dusty disk poses several questions for current models of ISM in spirals: how did so much dust end up (or form) so far outside the optical disk? Is this common or is this particular galaxy an anomaly? Is this pair truly a chance occultation or interacting in some way? 

To address these questions, we present here VLT/VIMOS-IFU observations of this pair with the specific aim to obtain the redshift difference between both galaxies, as well as a measurement of the attenuation curves in a distant spiral disk. 
This paper is organized as follows: section \ref{s:obs} describes the VIMOS observations, sections \ref{s:z} and \ref{s:sf} discuss the observed redshifts and emission line class of the foreground galaxy respectively, section \ref{s:obsfx} discusses two observational effects in determining the observed attenuation curve, section \ref{s:analysis} present our analysis of the VIMOS fibres, and section \ref{s:disc} is our discussion of these results.

\section{VLT/VIMOS Observations}
\label{s:obs}

The VIsible MultiObject Spectrograph \citep[VIMOS][]{Le-Fevre03a} on the VLT/UT3 can operate in three different modes; either as an imager, a Multi-Object Spectrograph, or as an Integral Field Unit (IFU), which is the mode used for this study.
The IFU consists of 4 quadrants of 1600 fibres each, feeding four different 2k$\times$4k CDDs. Each quadrant is made by four sets (pseudo-slits) of 400 fibres. 
We asked for $\rm 0\farcs67/fibre$ (27"x 27" field-of-view) with arcsecond seeing or better, to resolve as much detail of the overlap region as possible while covering the entire object. The medium spectral range mode (MR) is 4900-10150 \AA \ with R=720 spectral resolution with the {\em G475} filter. 

The galaxy pair is listed in the 2MASS \citep{2MASS} catalog under 2MASXJ00482185-2507365 (RA=00:48:21.8, DEC=-25:07:37) with a major axis diameter of $0\farcm5$.  During ESO period 84A (on 2009-12-18), 1.2 hours of service mode observations with VLT/VIMOS in IFU/MR mode were obtained  of the target pair (program 384.B-0059A). Seeing measurements were reported between $\sim0\farcs7-0\farcs8$, and the airmass was less than 1.02. 
The pointing successfully covered most of the pair (Figure \ref{f:fov}). 

Our observing strategy was based on a very tentative identification of the foreground galaxy's redshift as z=0.03 based on an HIPASS \citep{hipass1} radio spectrum. Thus, we opted for medium spectral resolution to maximize throughput and obtain the best signal-to-noise on the eventual attenuation curve. 
The spectral resolution of the MR spectral mode is not high enough to obtain kinematic information for either galaxy.
 
The VIMOS pipeline delivers high-level science products for observations in service mode. We use the flux- and wavelength-calibrated images, and the extracted fibre spectra\footnote{see also \url{http://ifs.wikidot.com/vimos} \citep{Westmoquette09} and the ESO's website \url{http://www.eso.org/}.}. 
The standard pipeline performs bias subtraction and extracts fibre-spectra, after solving for wavelength and dispersion using sky-lines, and resampling them to a constant wavelength step per pixel (2.6 \AA\ per pixel).
Individual spectra are corrected for fibre transmission and re-ordered in a spectral image for each quadrant with each fibre-spectrum in a row. The corresponding sky positions can be found in the position table {\em ifutablesHR.tar} (from the ESO website), which we combined into the field-of-view image in Figure \ref{f:fov}. 
We assigned WCS sky coordinates to this fibre map based on the assumption that the position of the brightest fibres correspond to the nuclei of both galaxies.
Small-scale corrections such as pixel-to-pixel response ("flat-fielding") and fringing correction are not executed in the VIMOS pipeline \citep{Izzo04} or subsequent improvements  \citep{Scodeggio05}. In addition to the pipeline reduction we flat-fielded and subtracted the sky emission as discussed below.

\subsection{Flat Fielding}
\label{s:flat}

As we are interested in the extinction curve by taking the ratio between fibres, the fibre spectra would not need be flat-fielded, provided the fibre response corrections are identical for both fibres. However, this is certainly not the case in VIMOS. Small-scale corrections add both noise and complicate the fit to the slope of the extinction curve.
Flatfielding is done on an individual fibre basis using the quadrant-specific flat-field files (MXFF). The lamp spectrum for a fibre is retrieved, and a third order polynomial is fit and divided out  to remove the lamp's black body spectrum. This normalized residual is the response function for flat-fielding and the spectrum is divided by this function. 

\cite{Zanichelli05} present a VIMOS graphical pipeline and point out that to correct for overall fibre response, one can use the emission in the 5577 \AA\ sky line, which should be uniform across the field-of-view. We checked using this line and the relative total fibre responses are uniform within a few percent.

\subsection{Sky Estimate and Subtraction}
\label{s:skysub}

After flat-fielding of each fibre as explained in section \ref{s:flat}, the nigh sky spectrum needs to be subtracted.
\cite{Zanichelli05} point out that the VIMOS instrument does not have dedicated fibres to estimate the sky background. However, in cases where the objects do not cover the entire field-of view, such as ours, they advocate the use of the median of those fibres with lowest total intensity as a good measure for the sky background. This is what we have done for each quadrant; we computed the average of the group of fibres in each quadrant, which are in the lower range of the summed fibre flux values.

\begin{table}
\caption{The identified lines in the spectrum of the foreground galaxy's nucleus (blue circle and spectrum in Figure \ref{f:fov}).}
\begin{center}
\begin{tabular}{l l l l l l l}
Line        		& $\lambda_{\rm em}$     & $\lambda_{\rm obs}$ & FWHM & EW & z & Line\\
			& 			& 			& 	&	&	& Strength \\
			& (\AA)			& (\AA)			& (\AA)	&	&	& (counts) \\
\hline
\hline

$\rm [HeI]$		& 5876 	 & 6380.6 & 11.75 &  2.10 & 0.0859 &  0.15 \\ 
$\rm [OI]$			& 6300 	 & 6670.5 &  5.09 &  1.64 & 0.0588 &  0.10 \\ 
$\rm [NII]$			& 6548 	 & 7182.5 &  4.71 &  4.22 & 0.0671 &  0.22 \\ 
$\rm H\alpha$ 	& 6563 	 & 6987.4 &  4.75 &  5.73 & 0.0647 &  0.33 \\ 
$\rm [NII]  $      		& 6583 	 & 6968.1 &  4.44 & -0.09 & 0.0642 & -0.01 \\ 
$\rm [SII]$        		& 6731 	 & 7008.4 &  6.33 &  4.62 & 0.0646 &  0.24 \\ 

\hline
\end{tabular}
\end{center}
\label{t:lines}
\end{table}%

\section{Interacting or Not? The Redshifts of Both Pair Members}
\label{s:z}

We estimate the redshifts of both galaxies based on the fibre spectra of their respective nuclei
(the blue and red circles in Figure \ref{f:fov}, respectively). Figure \ref{f:fov} also shows the usable spectrum from these fibres (4900-7300 \AA).

We identified five emission lines in the foreground galaxy's spectrum (although the [He] redshift identification is uncertain, Figure \ref{f:fov}). 
We fit a Gaussian profile to each line and used the central wavelength to estimate a redshift for the foreground galaxy (Table \ref{t:lines}).
The mean redshift is $z=0.0639 \pm 0.0027$ for the foreground galaxy (excluding the [He] line). 

Two absorption feature redshifts for the combined object (2MASXJ00482185-2507365) are reported in the literature; 
z=0.063641$\pm$0.0005 \citep{Ratcliffe98} and z=0.064011$\pm$0.000150 \citep[6dF survey,][]{Jones09}.
These are essentially redshift estimates for the background galaxy because it dominates the flux. 
Both previous studies identify the spectrum as that of a bulge or elliptical, dominated by 
old stellar spectra. 

In the spectrum of the background galaxy's nucleus (red circle and aperture in Figure \ref{f:fov}), we identified 
two absorption features (Mg and Na, indicated in Figure \ref{f:fov}). Their position, compared to template spectra 
for a Bulge or S0 \citep{Calzetti94,Kinney96}, confirm the previous redshift estimates ($z=0.06495 \pm 0.0007$).
In the outskirts of the background galaxy's disk, some faint star formation is apparent in the {\em HST} image. 
However, it is not strong enough for a certain emission line identification in the relevant fibre-spectra. 

The difference in redshift between the pair members is very small; $\rm \Delta z = 0.001 ~ (\Delta v_{\rm sys} = 300$ km/s), 
or only some 4 Mpc, assuming Hubble Flow ($\rm H_0=73 ~ \rm km/s/Mpc$). 
%

%
This separation is still substantial but the relative velocities do not completely preclude an interaction in this pair's history or future.
Indeed, both pair members could easily be at the same distance and hence strongly interacting. In our opinion, this is unlikely, because both galaxies appear otherwise undisturbed and the dusty structures appear to form a symmetric disk around the foreground spiral, not a tidal feature.

\section{Starburst or AGN? Line Strengths in the Foreground Dwarf Galaxy}
\label{s:sf}

If we compare the line strength in Table \ref{t:lines}, we find ratios of -0.18 and -0.14 for log([NII]/H$\alpha$) and log([SII]/H$\alpha$), respectively. Following the \cite{BPT,Kewley06} classification (BPT diagram), these line ratios identify the foreground galaxy as most likely an HII (star forming) object, not an AGN or LINER. If the foreground galaxy has an AGN, it is a fairly weak one (we lack H$\beta$ information to make a complete classification). If both galaxies are indeed in close proximity, the current star-formation in the foreground galaxy may well be tidally triggered.
\begin{figure*}
\begin{center}
\includegraphics[width=0.49\textwidth]{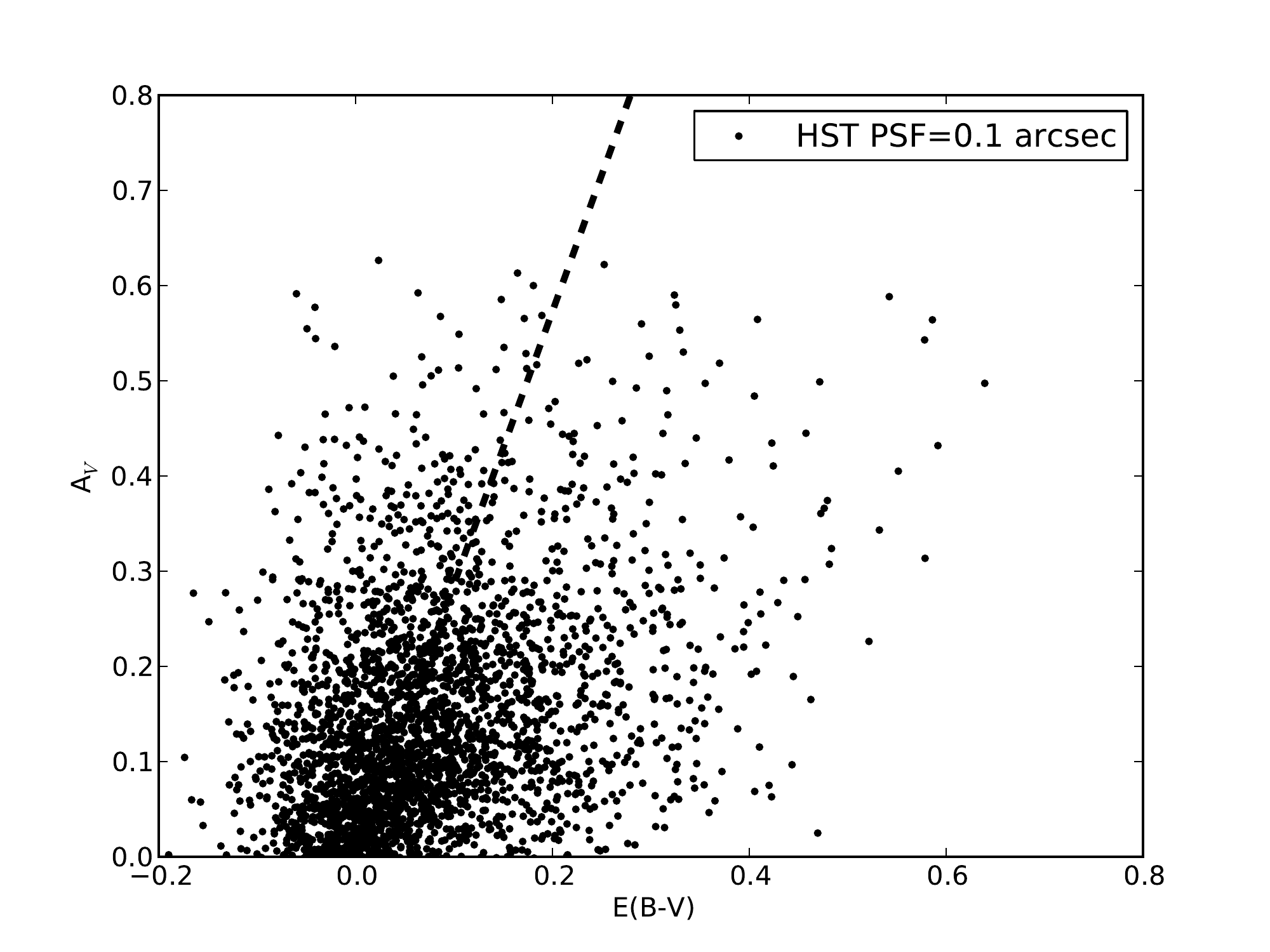}
\includegraphics[width=0.49\textwidth]{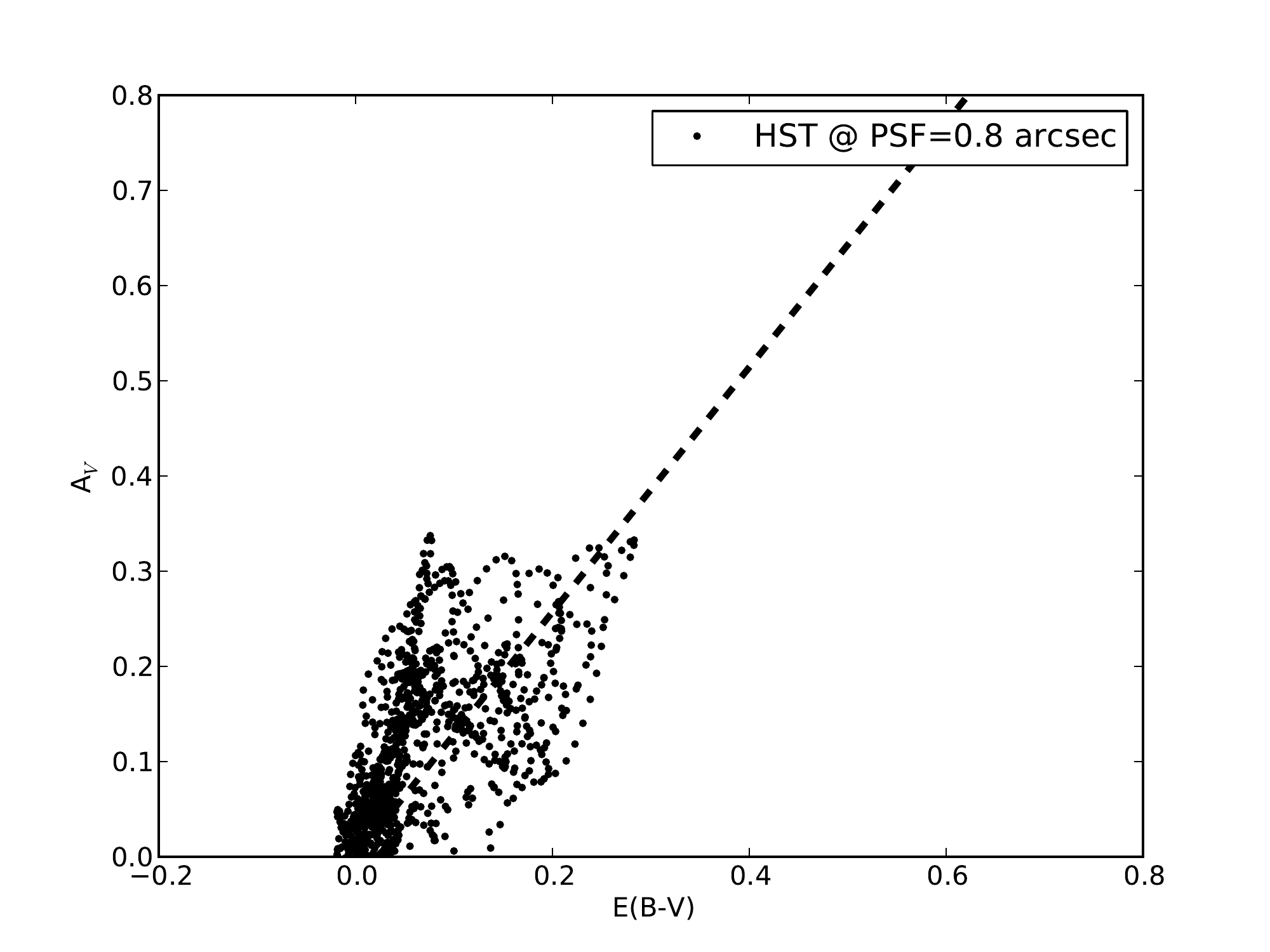}
\caption{The reddening, $E(B-V) = A_B - A_V$ versus attenuation in $F606W$, $A_V$ determined for each pixel in the attenuation map in Figure \ref{f:hst:Amaps}. 
For the optical depth maps at the native {\em HST} resolution a relation of $R_V = 2.86$ is retrieved (fit to the positive values of $A_V$). To compare with the VIMOS observations, we spatially smooth and resample the {\em HST} images to 0\farcs8 resolution with 0\farcs67 pixels  (right panel) to explore how the distance and seeing affect the extinction law that would be observed with `VIMOS resolution. Spatially smoothed and resampled images yield lower values of $R_V$, i.e., a grayer extinction law. The right panel shows the extinction law one would expect in the VIMOS observations based on the resampled {\em HST} images (FWHM=0\farcs8, pixelscale=0\farcs67); $R_V=1.29 $.}
\label{f:Rvd}
\end{center}
\end{figure*}

\section{Observational Effects on the Observed Extinction Curve}
\label{s:obsfx}

There are two observational effects to consider when comparing the observed relation between attenuation and wavelength in the overlap region: scattered light into the line-of-sight and the effects of physical sampling of the foreground ISM. Asymmetry in either galaxy introduces uncertainty in the extinction measures. Instrumental effects are discussed in section \ref{s:sky}.

\subsection{Scattering}
\label{s:scattering}

Because the galaxies are not a point sources but extended objects, the relation between attenuation and wavelength may include effects of light scattered {\em into} the line-of-sight by dust clouds in the foreground disk. This effect is treated in detail in the appendix of \cite{kw00a}. There are two sources of scattered light, the foreground and background galaxy. In Aperture I, the foreground galaxy contribution is removed (see below), and in Aperture II, this contribution is much lower than the background galaxy's. This leaves the background galaxy as the main source of scattered light in our measurement. Scatter would affect our measurement most where the background source is brightest, i.e., close to the background galaxy's nucleus, specifically Aperture II (see below).

\cite{kw00a} examine the contribution of scattered light into a line-of-sight as a function of angular and radial separation of the two galaxies. If the galaxies are sufficiently far apart, the scattering angle becomes small enough not to matter. They express the scattering percentage as a function of the ratio between radial extent and separation of both disks. The foreground galaxy disk has a radius of $\sim$12 kpc and the  separation implied by the respective redshifts is 4 Mpc. The ratio is thus several hundred which results in a scattering effect well under a percent of the measured flux. 

\subsection{Sampling}
\label{s:sampling}

In H09, we already noted the effect of spatial sampling on the observed extinction law. Assuming the background galaxy's redshift as an upper limit, we explored the effects of resampling the image on the attenuation measure. Now that the actual distance of the foreground disk is known, we can explore, using the {\em HST} attenuation maps as a start, predict the spatially averaged extinction curve that will be observed in the VIMOS spectra. 

Figure \ref{f:Rvd} shows the reddening ($E(B-V)$) versus attenuation ($A_V$) in the HST attenuation maps (Figures \ref{f:hst:fov} and \ref{f:hst:Amaps}) for the native resolution and the resampled case where the {\em HST} maps are spatially smoothed to the 0\farcs8 seeing and resampled at 0\farcs67 pixels of the VIMOS observations.
The spatially smoothed and resampled maps show a very gray observed extinction curve ($R_V << 3.1$), averaged over the whole HST overlap region at $R_V=1.3$. 
This implies that the notion of measuring the extinction curve as a function of position in the foreground disk of the foreground galaxy may be too ambitious for the actual sampling and resolution of the observations.

We adopt the term extinction curves and law for the remainder of the paper with the caveat that these measurements are influenced by sampling of the disk, with 
some scattered light from the background galaxy as a secondary effect in the very brightest sections of the background galaxy.

\subsection{Galaxy Asymmetry}
\label{s:asymmetry}

The central assumption in our analysis of the HST and VIMOS observations (below) is that both galaxies are rotationally symmetric. This assumption of symmetry must especially hold for the background galaxy. For this reason, we would prefer an elliptical galaxy as a background. However, the background galaxy in this case appears a very regular spiral without much structure from star-formation which often breaks symmetry on small scales. 
In principle, the dark dust structure visible in the HST image and the attenuation evident in the VIMOS data (Figures \ref{f:fov}, \ref{f:ccm:map} and \ref{f:hst:Amaps})
could be part of the background galaxy but it would uncharacteristically break symmetry for the background galaxy. And secondly, the general morphology of the dusty structures strongly suggests that they belong to the foreground disk (orientation, implied winding angle) and not the background galaxy.

\section{Observed Extinction Curves}
\label{s:analysis}

We analyze two parts of the overlap region: one where we have an estimate of the foreground galaxy's contribution (Aperture I in Figure \ref{f:fov}), and one where we assume there is little flux from the foreground galaxy (Aperture II in Figure \ref{f:fov}). Counterpart fibres for both foreground and background galaxies are retrieved from the blue and red or orange Apertures in Figure \ref{f:fov} respectively. 
We compute the attenuation as a function of wavelength from:
\begin{equation}
\label{eq:1}
A(\lambda) = -1.086 \times ln\left({[B(\lambda) e^{-\tau(\lambda)} + F(\lambda)] - [F'(\lambda)] \over [B'(\lambda)]}\right),
\end{equation}
\noindent where $B$ and $B'$ are the contributions from the background galaxy (overlap and counterpart respectively) and $F$ and $F'$ the contributions from the foreground galaxy (overlap and counterpart). The terms in square brackets correspond to an observable spectrum: overlap $[Be^{-\tau} + F]$, foreground $[F']$ and background $[B']$ counterparts. The implicit assumptions are that 
both galaxies are intrinsically rotationally symmetric and 
that the foreground counterpart fibre (F') has little flux contributed by the background galaxy due to the the exponential drop in background galaxy surface brightness with radius. As explained above, equation \ref{eq:1} ignores the effect of scatter and the effect of mixing transparent and opaque lines-of-sight.

\subsection{Aperture I: Foreground Stellar Disk}
\label{s:aperI}
\begin{figure*}
\begin{center}
\includegraphics[width=0.45\textwidth]{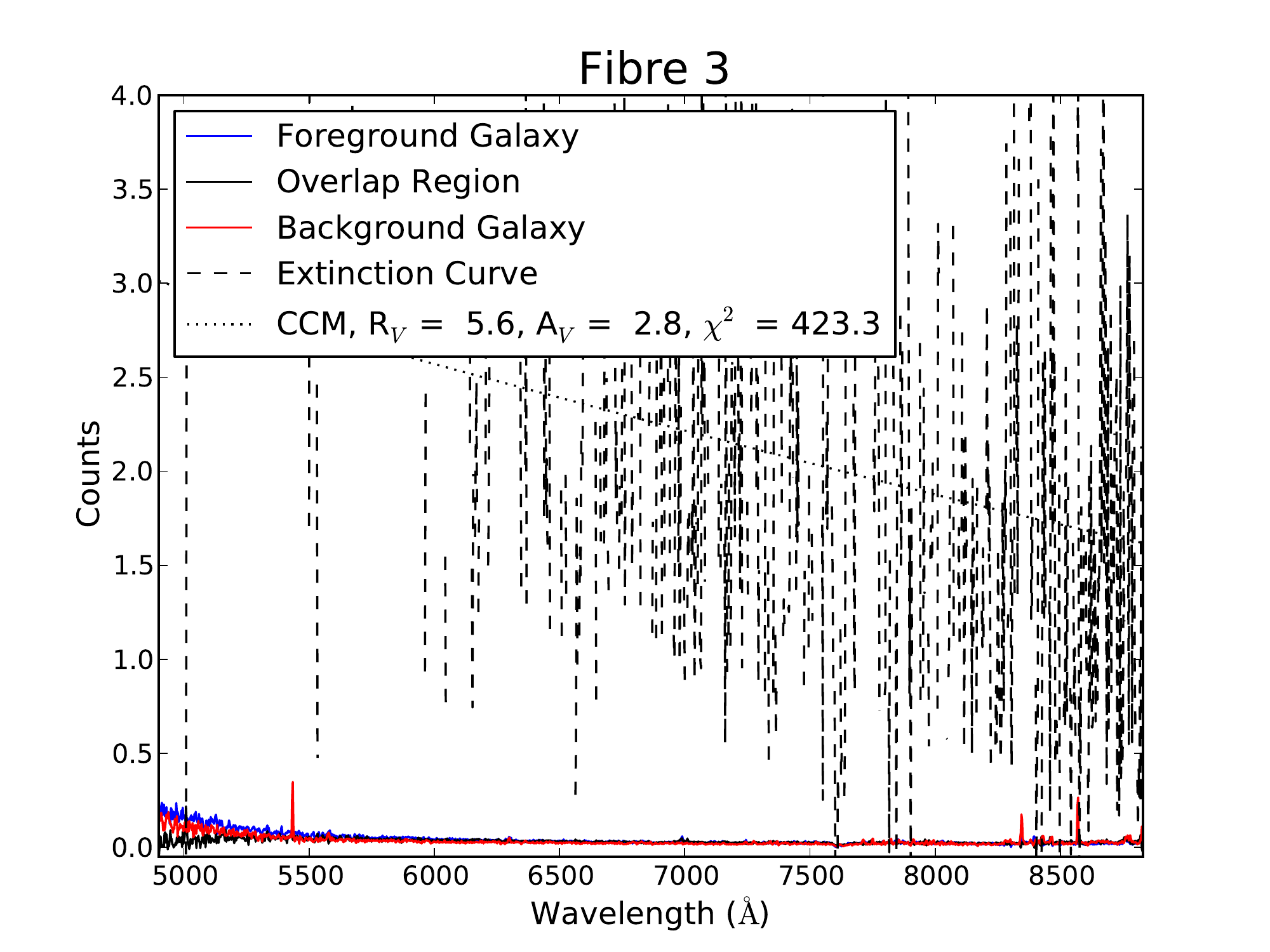}
\includegraphics[width=0.45\textwidth]{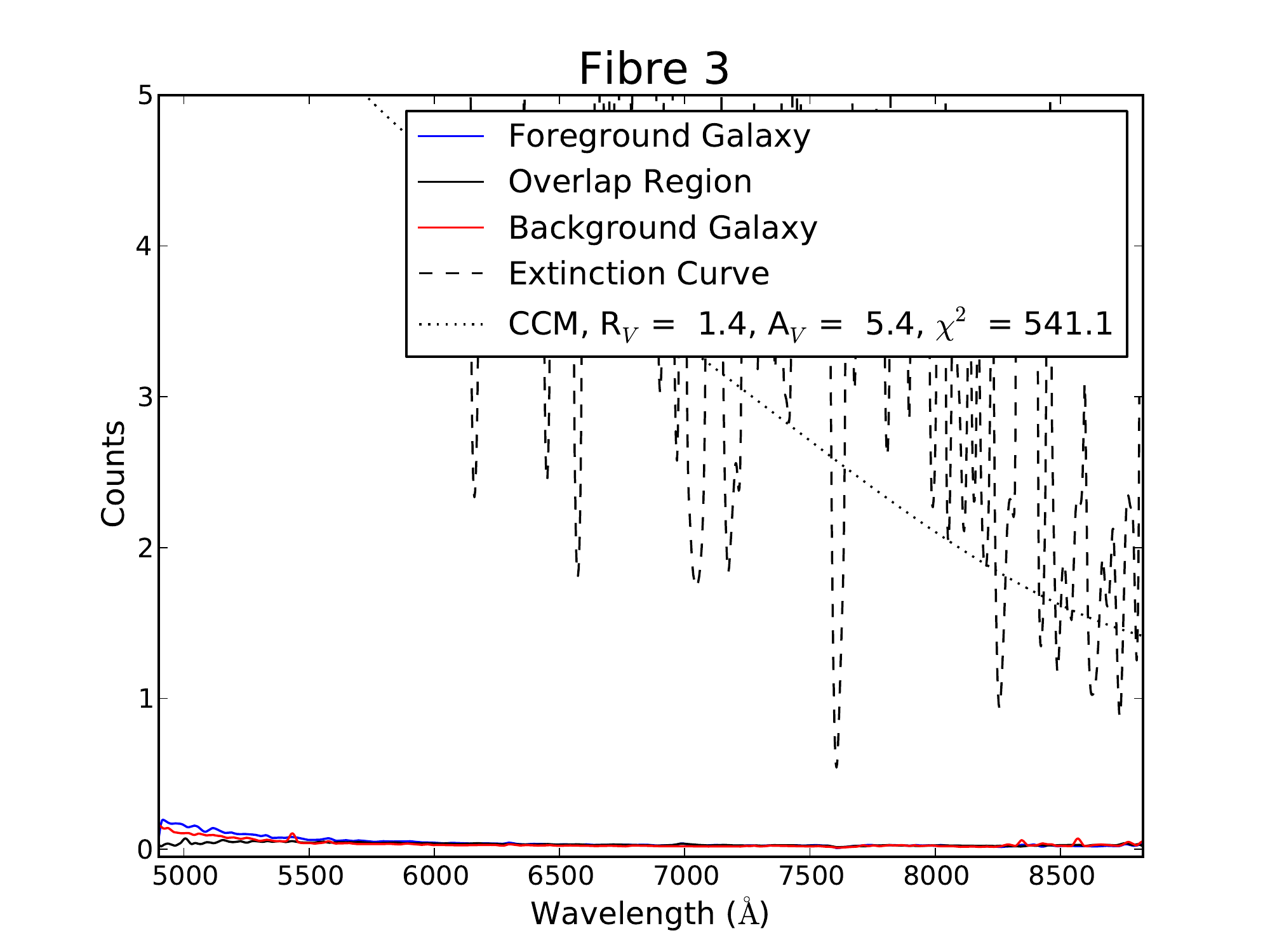}
\caption{An example of a raw (left) and spectrally smoothed (right) fibre spectrum in Aperture I.  The spectra were Hanning smoothed by 21 pixels (54.6\AA) to improve signal to noise. After aggressive spectral smoothing, a noisy extinction curve can be recovered. However, the CCM fit to this curve is problematic for both the slope and normalization. }
\label{f:aperI:f3}
\end{center}
\end{figure*}

In Aperture I (green box in Figure \ref{f:fov}), the flux contribution from the foreground galaxy ($F$) is not negligible and we use equation \ref{eq:1} to estimate the attenuation in the foreground disk ($[Be^{-\tau} + F]$, foreground $[F']$ and background $[B']$ are the green, blue and red box in Figure \ref{f:fov} respectively). 
%
%
The resulting extinction curves from the raw spectra are uncertain and noisy (Figure \ref{f:aperI:f3}), as the flux contribution from the foreground galaxy is first subtracted and the background galaxy is relatively faint, compounding the issue of individual pixel signal-to-noise. A viable extinction curve cannot be derived from the individual, raw spectra.

There are two ways to mitigate this, either by spectrally smoothing the individual fibre spectra, sacrificing spectral resolution, or by combining multiple fibres, sacrificing spatial information.
First we choose the first approach; Figure \ref{f:aperI:f3} shows the result of smoothing one of the fibres spectrally by 21 pixels (300 \AA), to improve the signal-to-noise in the extinction curve. However, only a very noisy extinction curve can be obtained. Even if we omit subtracting the foreground galaxy spectrum, no realistic extinction curve can be retrieved.

Thus, we can combine the spectra from the two least noisy fibres in Aperture I  into a single average extinction curve (Figure \ref{f:tau}, green line), but even this ``mean" extinction curve holds very little information.

\subsection{Aperture II: Foreground Outer Disk}
\label{s:aperII}
\begin{figure*}
\begin{center}
\includegraphics[width=0.49\textwidth]{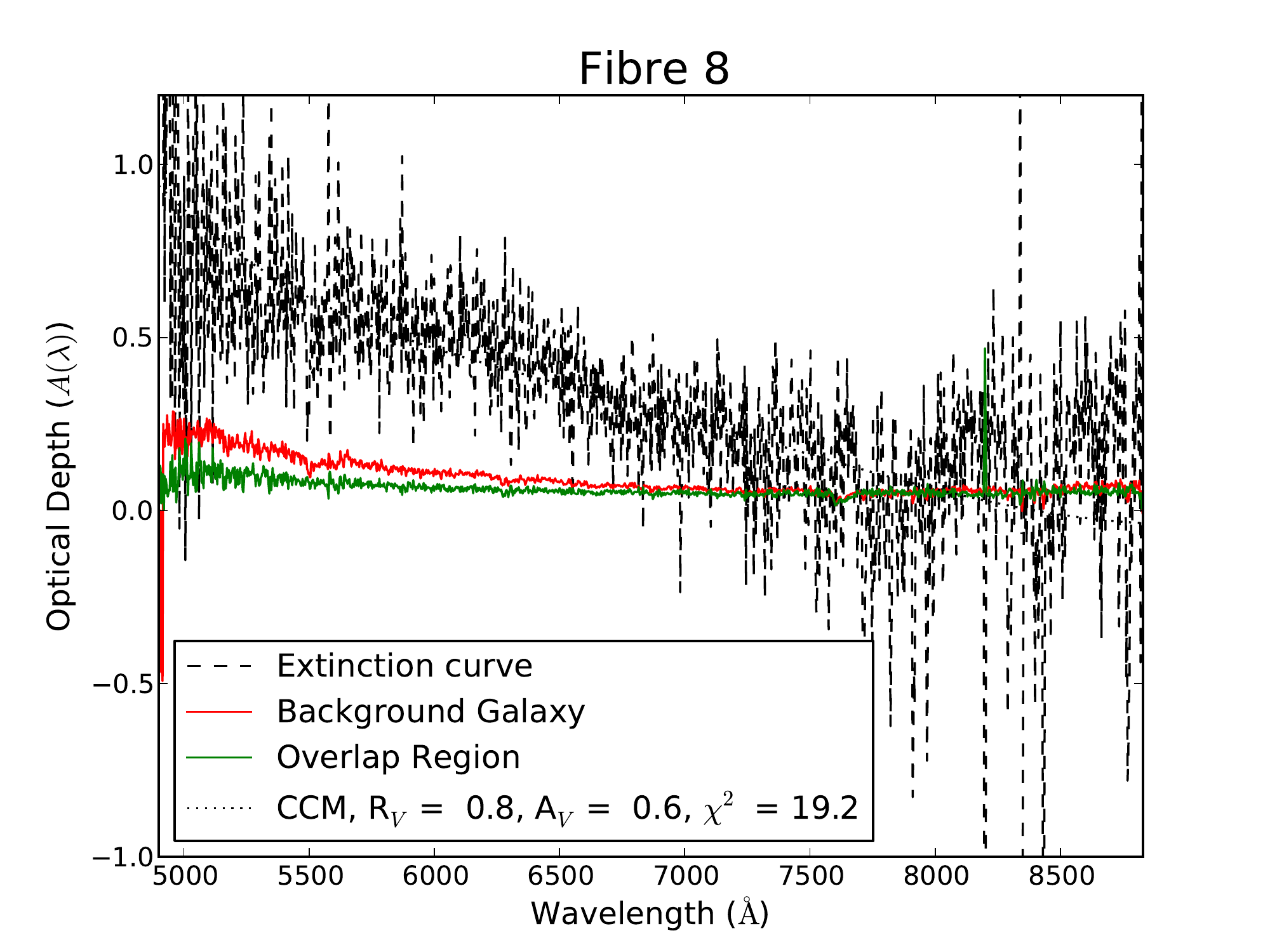}
\includegraphics[width=0.49\textwidth]{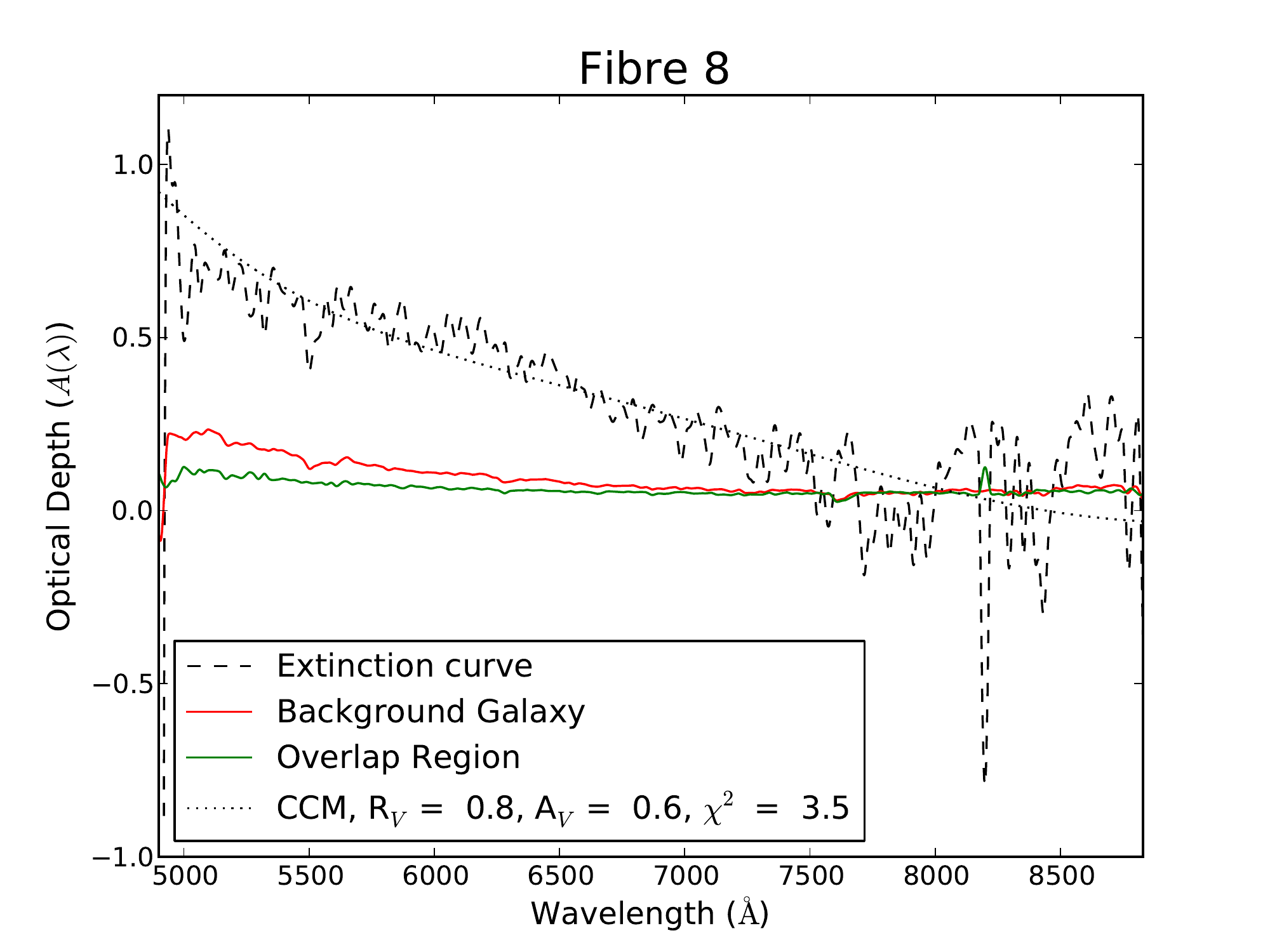}
\caption{An example of a raw (left) and spectrally smoothed (right) fibre spectrum in Aperture II. The spectrally smoothed spectrum was convolved with a 21 pixel Hanning profile. The CCM fits to both extinction curves agree well ($R_V=0.9, A_V=0.9$). The normalization is close to the values derived from the {\em HST} attenuation map in the $F606W$ filter (Figure \ref{f:hst:Amaps}). The fit to the slope ($R_V$) is lower than expected from the HST broad-band filters. }
\label{f:aperII:f7}
\end{center}
\end{figure*}

In Aperture II (black box in Figure \ref{f:fov}), our analysis is simpler, because we can ignore the flux contribution from the foreground galaxy, as the background galaxy dominates ($F<<B$). 
Aperture II encompasses the overlap region we analyzed in H09 (Figure \ref{f:hst:fov}). We estimate the optical depth as a function of wavelength in these fibres from:
\begin{equation}
\label{eq:2}
A(\lambda) = -1.086 \times ln \left({[B(\lambda)] \over [B'(\lambda)]}\right).
\end{equation}

Individual extinction curves for the 15 fibres of Aperture II are shown in Figure \ref{f:aper2:fib}. 
We note that in the majority of fibres in this Aperture, there is enough s/n even in the unsmoothed spectra to generate a (noisy) extinction curve. 
The one exception where a CCM fit was problematic (Table \ref{t:ccm} and Figure \ref{f:aper2:fib}), fibre F15, is the one closest to Aperture I (Figure \ref{f:aper2:fibrepos}), where it proved impractical to infer an extinction curve.
We can attempt to improve the s/n in the extinction curve in the same two ways as above: by spectrally smoothing individual fibre spectra in the spectral dimension and/or by combining spectra, in effect averaging spatially.
\begin{figure}
\begin{center}
\includegraphics[width=0.45\textwidth]{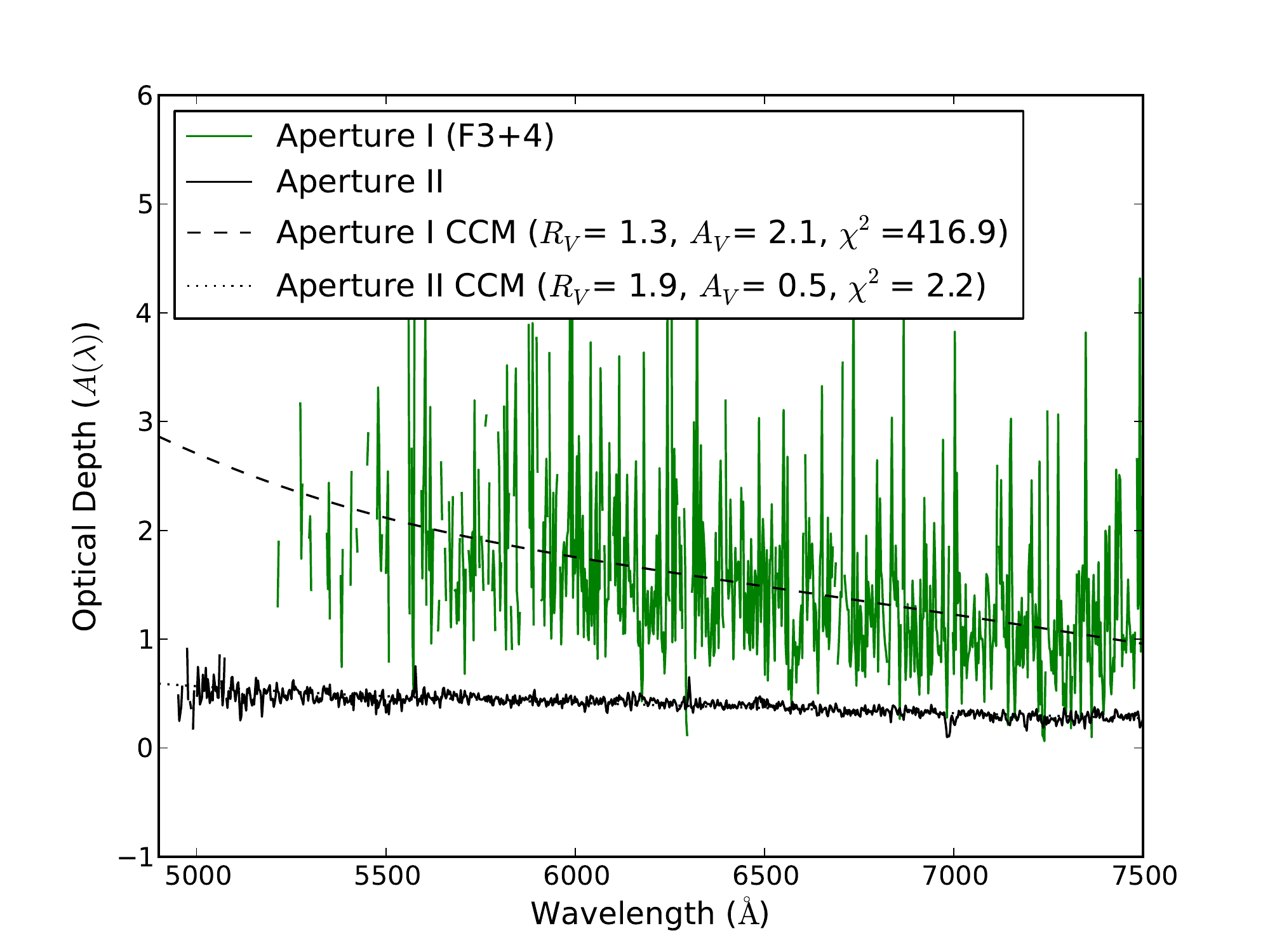}
\caption{The two mean extinction curves for both apertures; green line (the two best fibres in the green box in Figure \ref{f:fov}) and black line (Aperture II in Figure \ref{f:fov}). 
We show fits of the CCM relation to these two extinction curves. Aperture I shows extremely high $A_V$ values and an unexpectedly high value of $R_V$. The mean values for Aperture II are in line with expectations from the {\em HST} aperture, both in normalization ($A_V$) and slope ($R_V$).}
\label{f:tau}
\end{center}
\end{figure}

Figure \ref{f:aper2:smoothing} illustrates the effect of spectral smoothing an individual fibre spectrum. While the noise is reduced, the fit to the extinction curve (see below) remains the same (but with lower $\chi^2$). 
The mean extinction curve from the averaged spectra in all of Aperture II (and the corresponding background galaxy aperture) is shown in Figure \ref{f:tau}. 
By sacrificing the spatial information, the noise can also be  brought down. One thing is immediately evident: the spatially averaged extinction curve is very flat, i.e., a gray extinction curve, in line with what one would expect for an average of the whole aperture from \S \ref{s:sampling}.

\subsection{Uncertainties and Biases}
\label{s:sky}

In section \ref{s:obs}, we point out that we supplement the VIMOS pipeline with sky-subtraction and a fibre small-scale sensitivity correction. In the analysis above, we compare the fluxes different fibres to each other. Differences in the fibre-to-fibre transmission, sky-subtraction or wavelength response (flat-fielding) could affect the resulting fibre extinction curves, both the normalization ($A_V$) and the slope ($R_V$).

The spectral response for individual fibres can affect both the slope ($R_V$) and normalization ($A_V$) of the derived extinction curve. The relative correction for overall fibre response, done in the pipeline, for both the fibre in the overlap region and the comparison one(s) in either galaxy affects the normalization of the extinction curve. 
In addition, the small scale, pixel-to-pixel correction affect the overall fit. For this reason, we supplemented the pipeline with individual, normalized, fibre flat-fields.
Flat-fielding predominantly changes the noise and spectral slope of the fibre spectra. 
If both the fibre in the overlap (either aperture) and the background galaxy fibre were affected in the same manner, 
it would only make the CCM fit somewhat noisier but there are subtle differences in the slopes of the VIMOS fibre flat-fields, which could cause a bias in the slope, $R_V$ as well.

Sky-subtraction has the potential to change both the normalization and slope of the extinction curve but only if there are spatial variations in the sky background. The field-of-view of the VIMOS/IFU is sufficiently small that one would not expect any real variation in the sky emission, a fact we used to check the individual fibre transmission.
However, each quadrant of the IFU has a slightly different throughput in the optics and for this reason we compute sky estimates for each quadrant. 

Fringing is most evident in the red part of these VIMOS spectra (the VIMOS received an upgrade in 2010 to counteract this feature as well as flexure). In principle, this does not affect our extinction curve estimate except that it makes the spectra and resulting extinction curve more noisy.

\begin{figure}
\begin{center}
\includegraphics[width=0.49\textwidth]{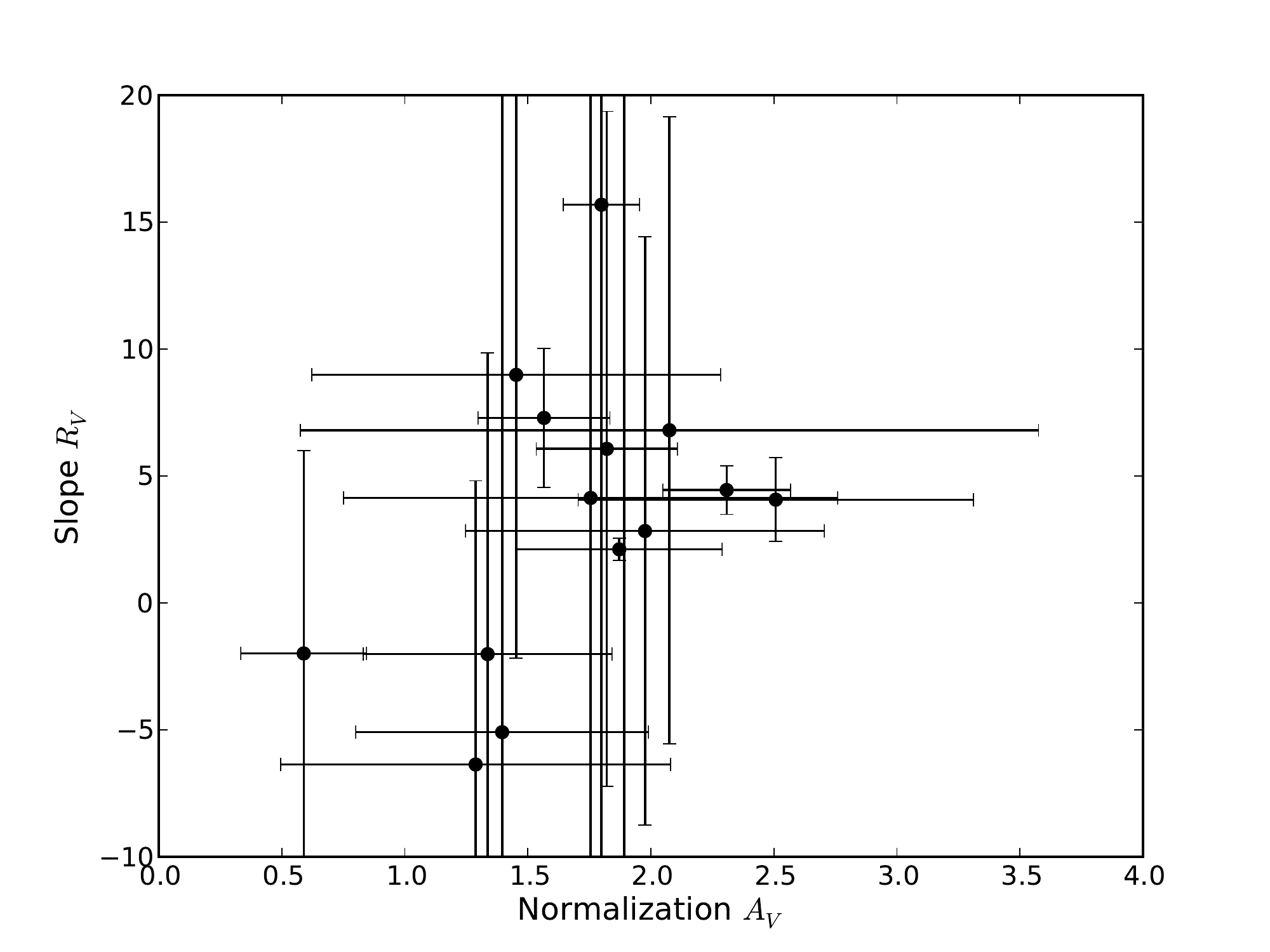}
\caption{The $A_V$ and $R_V$ values with the spread due to different sky-subtraction. Each quadrant's sky estimate was subtracted in turn from the overlap and comparison spectrum and the rms of the resulting CCM fits is shown as the uncertainty. }
\label{f:ccm:err}
\end{center}
\end{figure}

Differences in the sky-subtraction or transmission between fibres would affect the resulting extinction curve. This relative response calibration is well taken care of in the VIMOS pipeline with the relative transmission file. 
\cite{Zanichelli05} point out that the VIMOS instrument does not have dedicated fibres to estimate the sky background. However, in cases where the objects do not cover the entire field-of view they advocate the use of the median some fibres with low total intensity as a good measure for the sky background. This is what we have done for all our fibre spectra; subtracted an appropriate sky background for each IFU quadrant (e.g., Q4 in Figure \ref{f:fov}). 
To estimate the effect of sky subtraction, we repeated the CCM fit with the sky estimate for each quadrant subtracted from the overlap fibre and the corresponding background fibre in Aperture II. The rms spread in both values, the normalization and slope is shown in Figure \ref{f:ccm:err}. 
Typical rms values in the normalization are $A_V < 1$ but with larger uncertainties in the slope, $R_V$.

\begin{figure*}
\begin{center}
\includegraphics[width=0.49\textwidth]{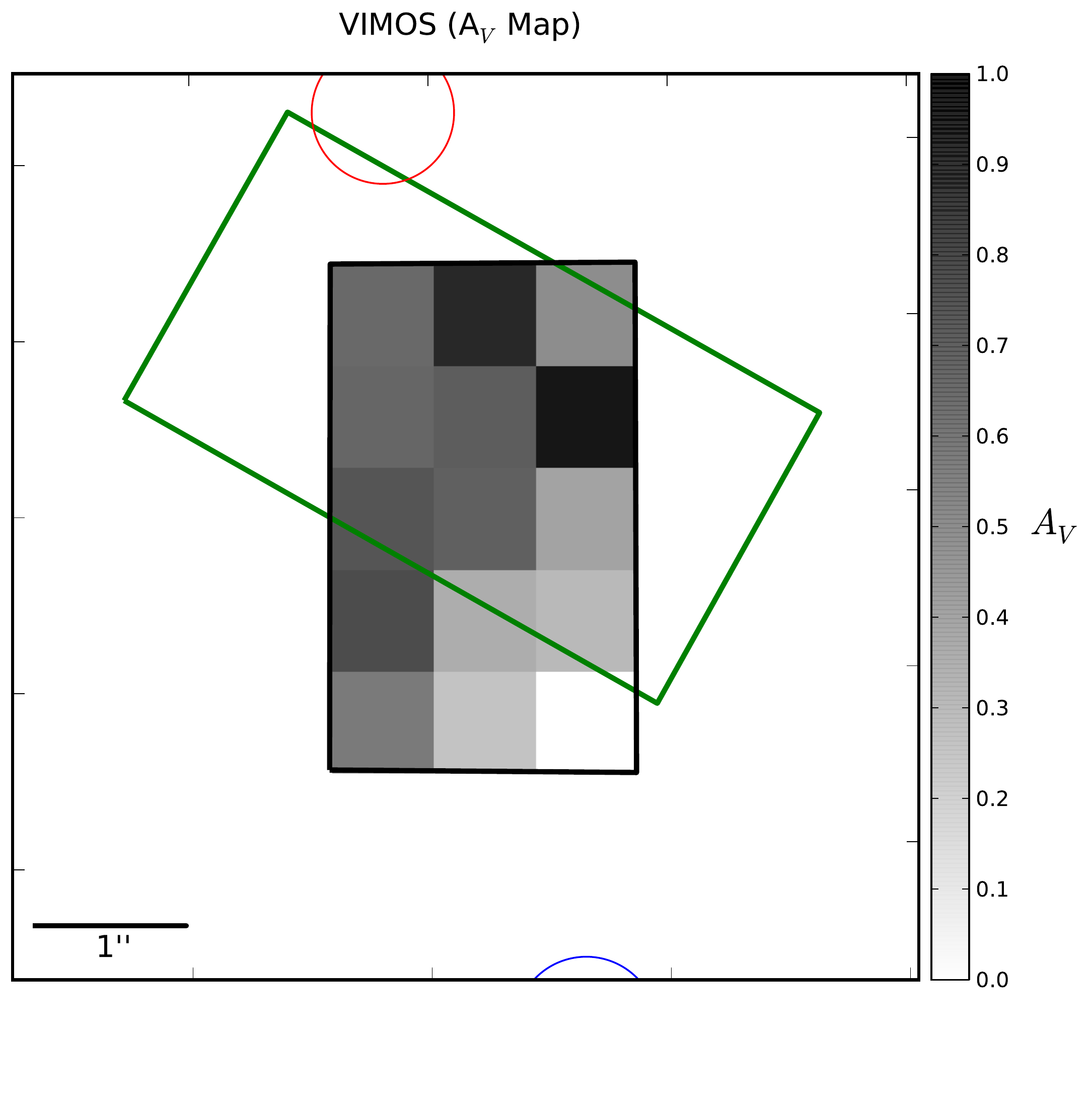}
\includegraphics[width=0.49\textwidth]{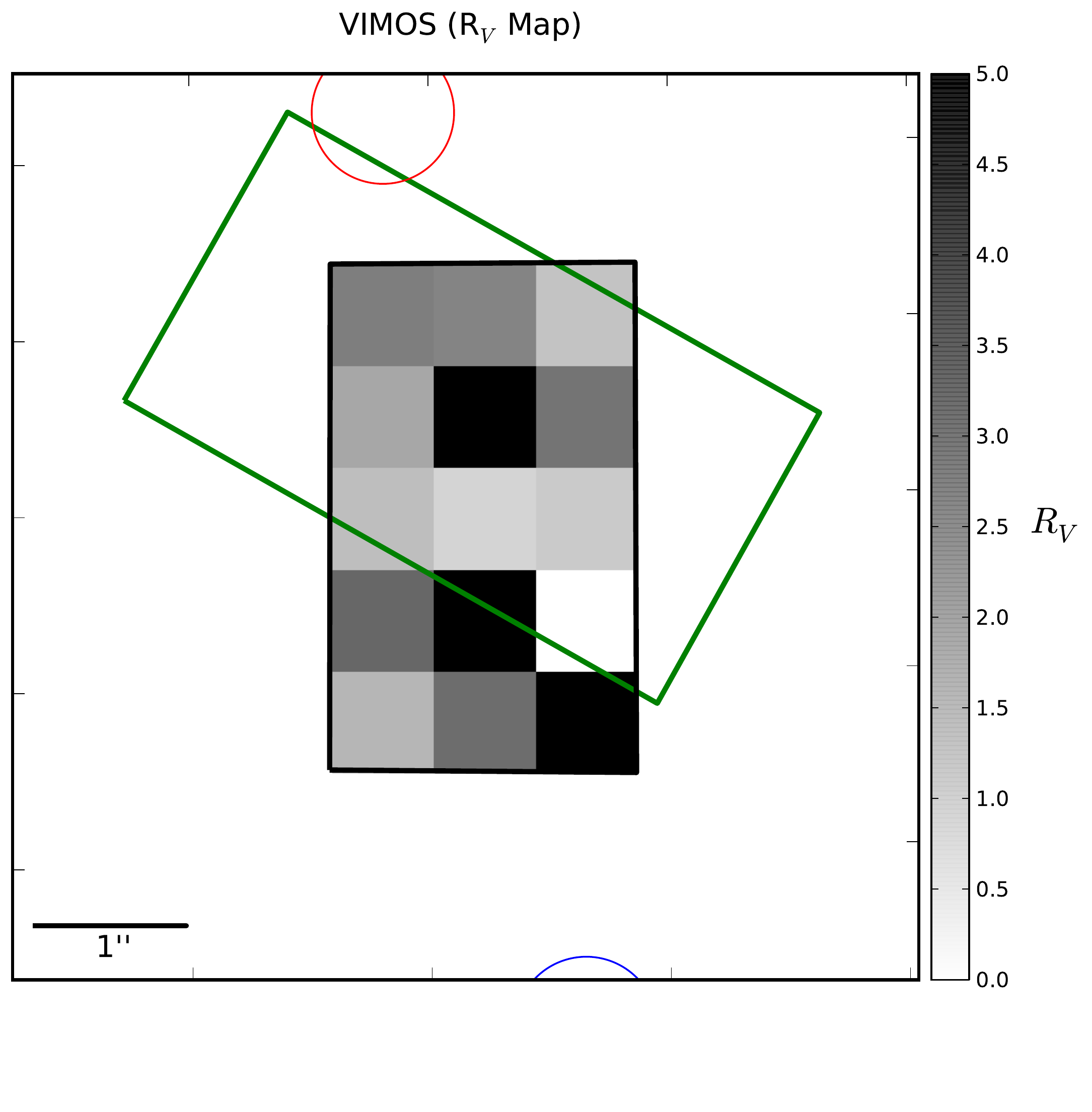}
\caption{Maps of the CCM fits, the slope $R_V$ (left) and normalization, $A_V$ (right). The black rectangle is the aperture analyzed in H09 (Figures \ref{f:hst:fov} and \ref{f:hst:Amaps}). The red and blue circles denote the background and foreground galaxy nuclei, respectively to help with orientation.
The VIMOS observations cover the dark spiral structures observed in the {\em HST} extinction map. The $A_V$ values in both maps agree well. 
Realistic $R_V$ values range from 0.8 to 5.2, on average much lower than the Milky Way value ($R_V=3.1$).}
\label{f:ccm:map}
\end{center}
\end{figure*}
\begin{figure}
\begin{center}
\includegraphics[width=0.5\textwidth]{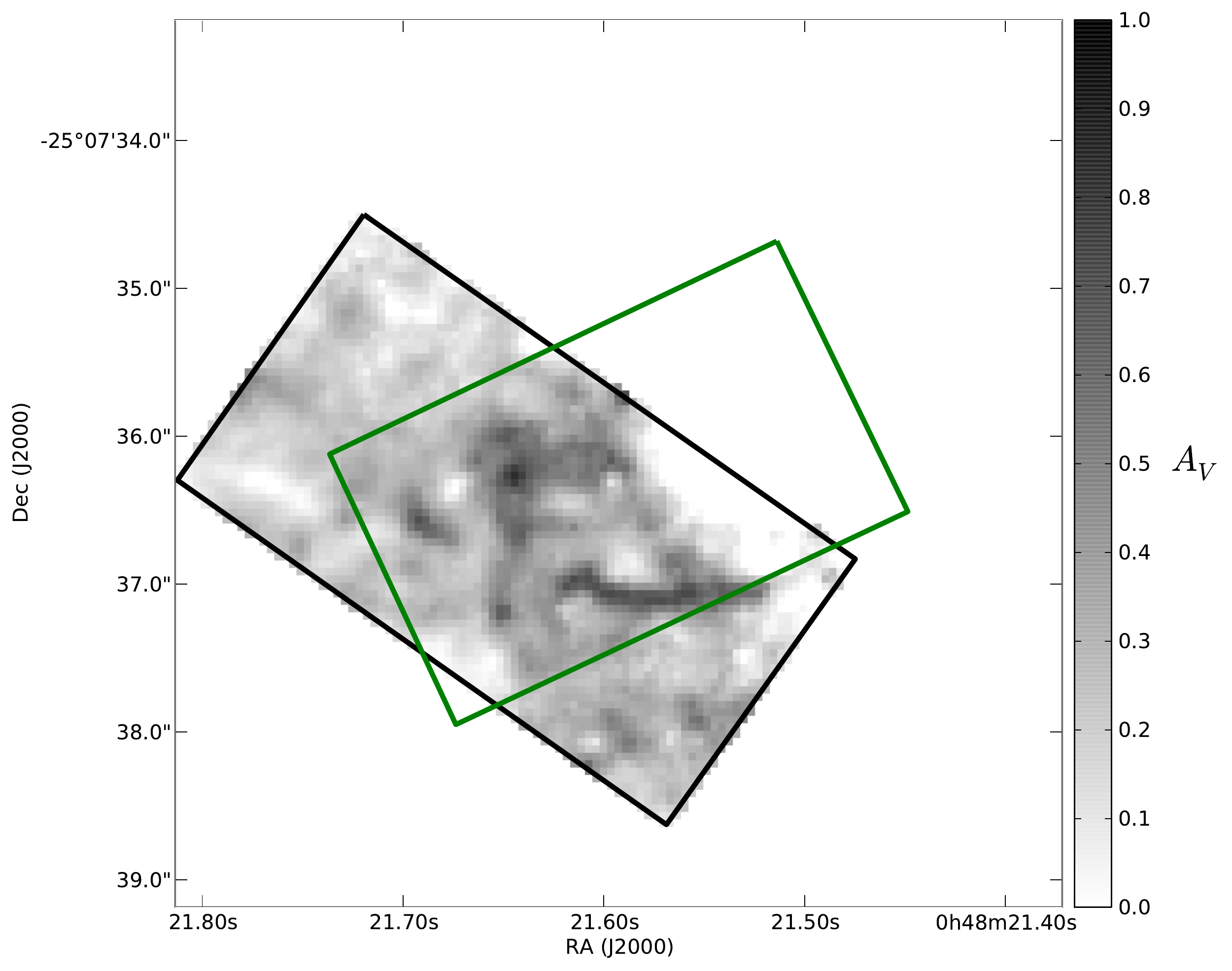}
\caption{The extinction ($A_V$) map based on the $F606W$ {\em HST} image (black outline), using method B described in H09 (using a model background galaxy). Aperture II in the VIMOS analysis is marked with the green triangle. }
\label{f:hst:fov}
\end{center}
\end{figure}
\begin{figure}
\begin{center}
\includegraphics[width=0.5\textwidth]{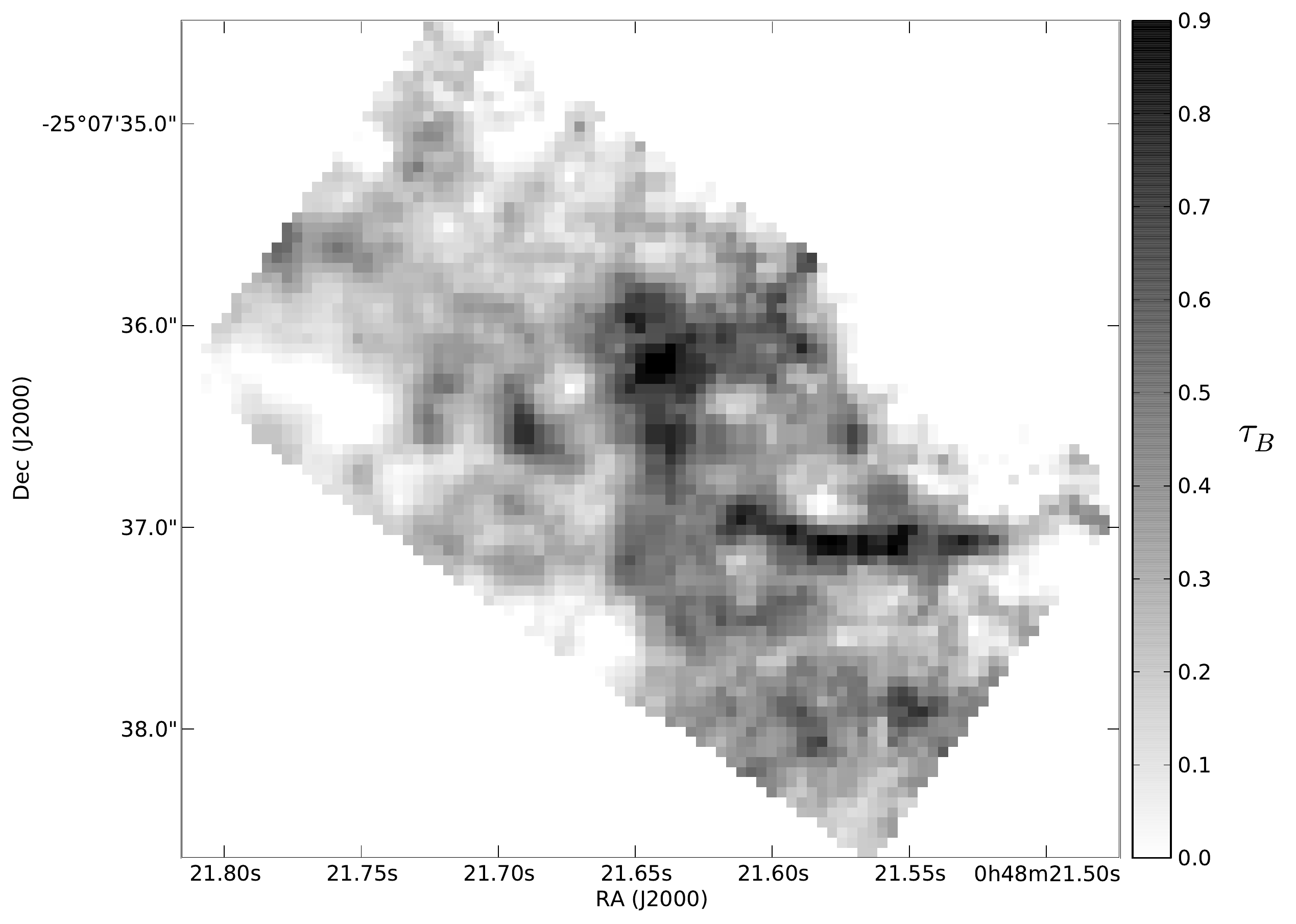}
\includegraphics[width=0.5\textwidth]{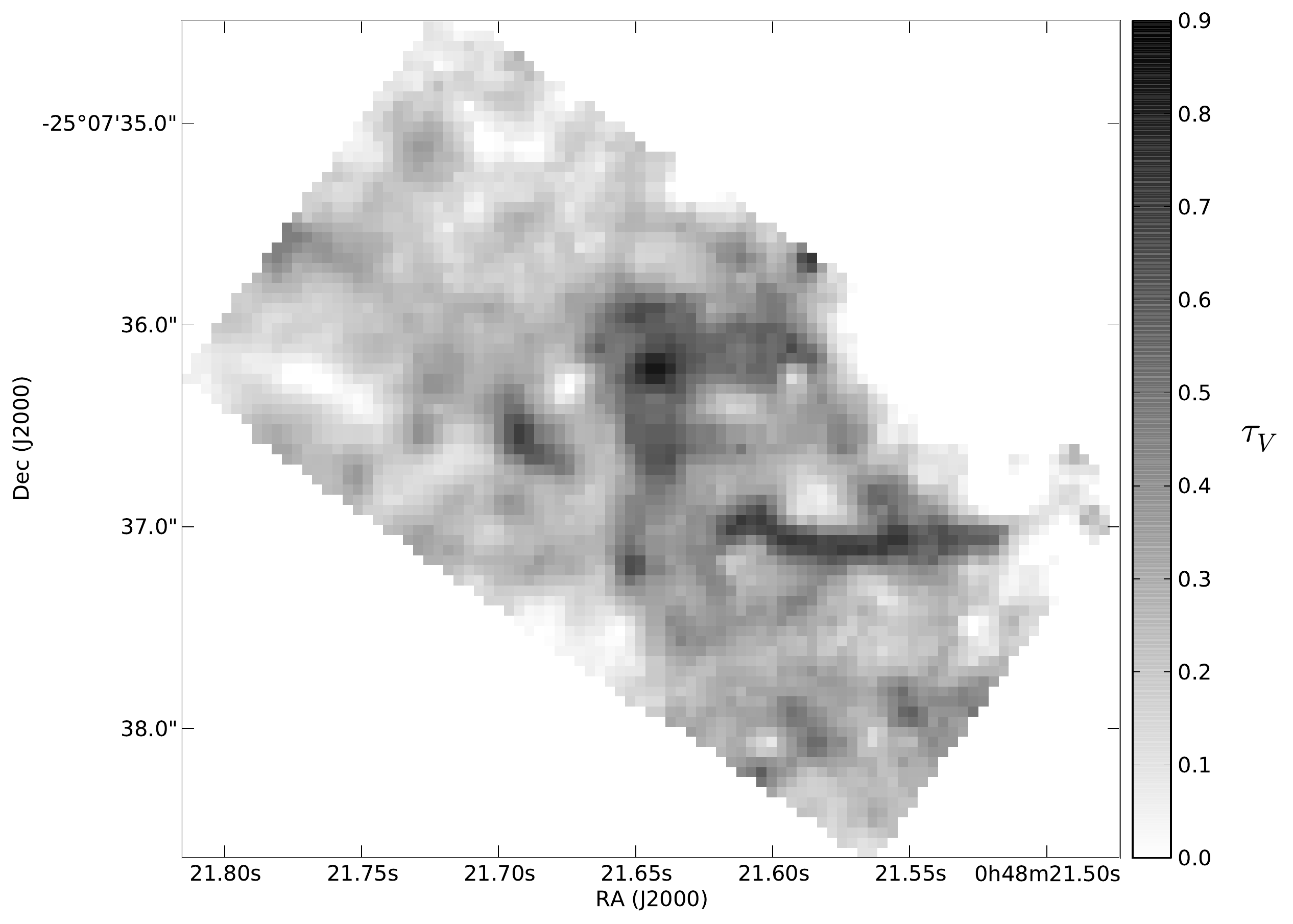}
\includegraphics[width=0.5\textwidth]{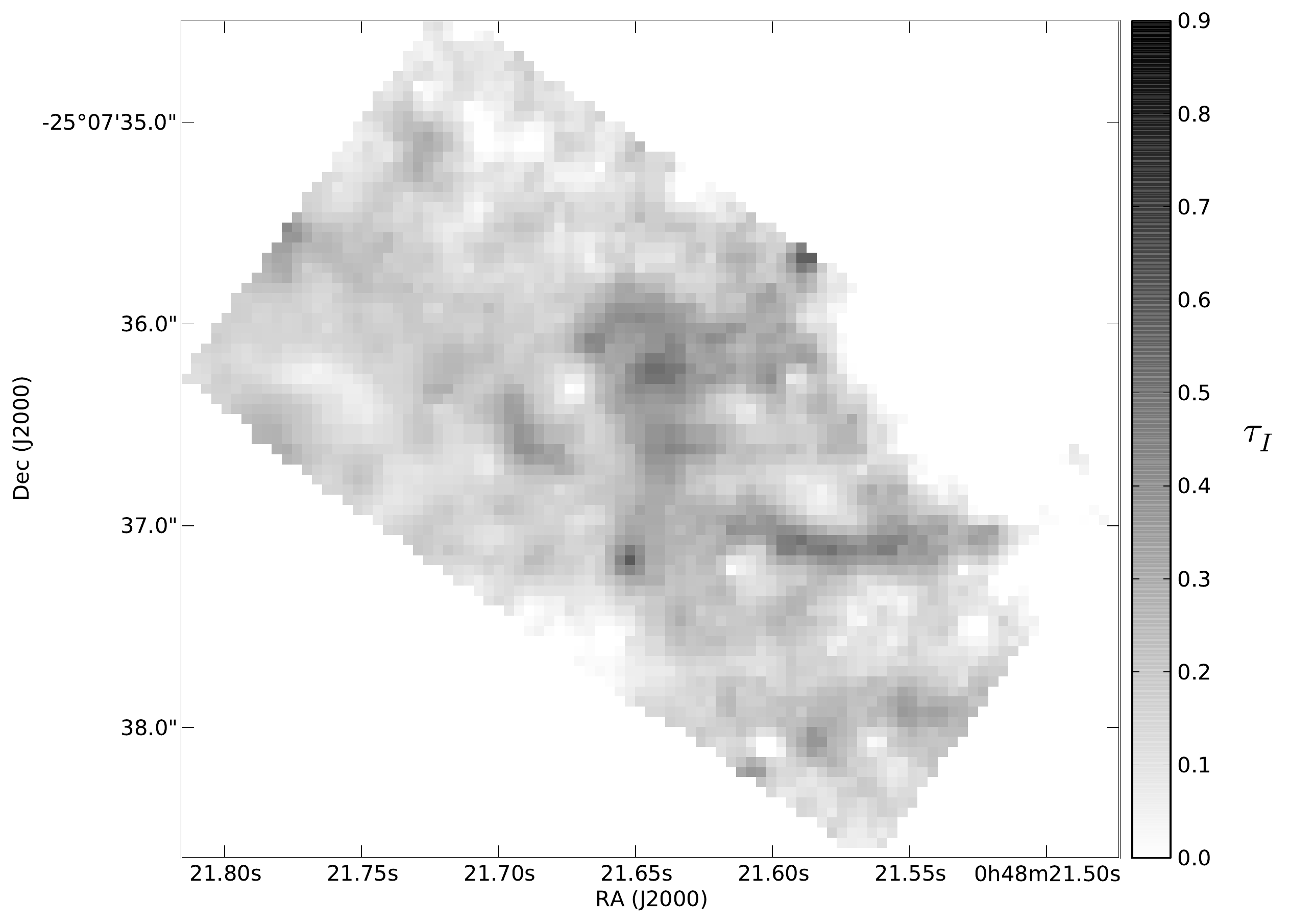}
\caption{Optical depth maps ($A_V = 1.086 \times \tau_V$) based on the three HST filters (F475W (B), F606W (V), and F814W (I). Optically thick structures stand out in the shorter wavelengths.}
\label{f:hst:Amaps}
\end{center}
\end{figure}

\subsection{Mapping the Extinction Curve}
\label{s:ccm}

We fit the Cardelli, Clayton and Mathis (CCM) extinction curve to each of the fibre extinction curves (really attenuation curves, see section \ref{s:obsfx}) in Aperture II. 
Figure \ref{f:aperII:f7} illustrates the effects of spectrally smoothing the fibre-spectra; the fit improves but generally the results did 
not change for each fibre (see also Figures \ref{f:ccm} and \ref{f:aper2:smoothing}). 
Only for the extreme value of $R_V$ is there a marked difference between the spectrally smoothed and the unsmoothed fit results.
We therefore map the fit results to the unsmoothed fibre extinction curves in Figure \ref{f:ccm:map} (also listed in Table \ref{t:ccm}).

The slope of the extinction curve ($R_V$) is between 1.5 and 4.2 in most realistic cases (Figure \ref{f:ccm:map}, right), with the majority of the values lower than 
the canonical $R_V = 3.1$ but often higher than we found by averaging the whole {\em HST} aperture (Figure \ref{f:hst:fov}). 
A plausible explanation for this is that the spatial averaging is not over the whole {\em HST} aperture but sections of 0\farcs67 squared.
Unphysical fits (negative $R_V$) are found in those fibres closer to the foreground disk (f14 and f15). The foreground disk contribution may not be non-negligible here (but we lack comparison fibres) and/or the spiral arm structure evident in the HST image may deviate from symmetry.
%
The grayer extinction curve seen in the overlap region (Figure \ref{f:ccm:map}) is an observational effect of the VLT fibre's coarse spatial sampling. 
The seeing and IFU fibre scale ($\sim 0\farcs7$), each aperture has a linear size of $\sim0.9$ kpc in the foreground disk.
At much higher spatial resolution, we found a Milky Way-like extinction law using the Hubble data with $\sim 60$ pc resolution elements (e.g., H09). 
In contrast, the VLT fibre apertures each cover a much wider area, each of which include both optically thin and thick regions of the foreground disk (Figure \ref{f:hst:fov}). 
The disk opacity is dominated by the optical thick regions whose covering factor does not change much with wavelength \citep[see also][for the extinction 
law effects of an inhomogeneous screen]{Natta84, Fischera05c}. This resolution effect has been observed in other occulting pairs as well 
\citep{Berlind97, kw01a,kw01b}. A Milky Way-type law occurs when the physical resolution is similar to the typical size of a Giant Molecular Cloud 
($\leq100$ pc), still an order of magnitude below the VIMOS sampling.

The $A_V$ values (Figure \ref{f:ccm:map}, left) are in line with averages of the extinction maps presented in H09. 
Figure \ref{f:hst:fov} shows the HST aperture with Aperture II overlaid. We note that the opaque structures in the {\em HST} 
extinction map (based on $F606W$ filter) dominate in Aperture II (e.g., the high value of $A_V$ in fibres 6, 8, and 12).

The fact that we still measure a slope in these fibres point to structures with similar extinction properties larger than what we found before ($>60-100$ pc), but smaller than the VIMOS sampling ($<9$ kpc), since $R_V$ values are still lowered by oversampling. The spiral structures in the {\em HST}-based extinction map (Figure  \ref{f:hst:fov}) are a prime candidate for these structures.

An alternative explanation is that the typical $R_V$ values of this galaxy's dusty ISM are higher than the canonical $R_V$ to begin with (before resampling). In this case, the grain distribution responsible for the extinction curve would be distributed more towards {\em smaller} grains (more reddening), a sign of recently reprocessed or produced dust. If this were the case, this would point to a disk undergoing significant change, for example through the star-burst and shocks associated with a close encounter.

\begin{table}
\begin{center}
\begin{tabular}{l l l l}
Measure	& Value & Unit		& Note \\
\hline
\hline

z			& 0.064 		& 		& \\
D 			& 263 		& Mpc 	&\\
m - M 		& 37.1 		& 		& \\

$M_{\rm F475W}$ 	& -18.2 		& mag.	& (a) \\
$M_{\rm F606W}$ 	&-18.9 		& mag.	& (a)\\
$M_{\rm F814W}$	& -19.5		& mag.	& (a)\\

B-V			& 0.64		&		& \\
V-I			& 0.62		& 		& \\
\hline


$log(M/L_V)$ 	& 0.26	& 	$L_\odot/M_\odot$ 	& (b)\\ 
$log(M/L_V)$ 	& -0.6	& 	$L_\odot/M_\odot$	& (c)\\ 

$M_*$	& $5.5 \times 10^9$		& $M_\odot$ & (b) \\
$M_*$	& $0.76 \times 10^9$	& $M_\odot$ & (c)\\

$<M_*>$	& $3.13 \times 10^9$	& $M_\odot$ & (d)\\

\hline

$<\tau>$		& 0.122		& 				& (d)\\				

$\Omega$	& 129.7		& arcsec$^2$ 		& (e)\\

$M_{\rm dust}$	& $0.52  \times 10^6$		& $M_\odot$	& \\


Gas-to-dust	& $\sim$100-500	&				& (f)\\
$M_{\rm gas}$	& $ 0.52- 2.6 \times 10^8 $	& $M_\odot$	& \\		


$f_{\rm gas}$		& 0.02-- 0.08		&				& \\
\hline
\hline

\end{tabular}
\end{center}
\caption{Characteristics of the Foreground Galaxy.}
\begin{tablenotes}
\item (a) Measured from the {\em HST} image.
\item (b) \cite{BelldeJong}, Figure 9, [B-V] - $log(M/L_V)$.
\item (c) same but [V-I] - $log(M/L_V)$
\item (d) Mean optical depth per pixel in the overlap region in H09.
\item (e) Kron aperture angular size.
\item (f) Gas-to-dust ratios vary across spiral disks \citep{Smith10b, Leroy11, Magrini11b} and these high values are typical for smaller disks.
\end{tablenotes}

\label{t:fg}
\end{table}%

\section{Discussion}
\label{s:disc}

With a new, more reliable, distance measurement for the foreground galaxy, we can use the HST observations to infer some of its global characteristics. Our aims were to gauge if the foreground galaxy is in any way unusual, specifically if it is interacting or if it is a fairly typical dwarf spiral which just happens to be positioned over a bright neighbour. 

We first estimate the stellar mass of the foreground galaxy. 
We calculate a V band absolute magnitude of $M_V = -18.9$, based on a 9\farcs0 (11.5 kpc) elliptical (Kron) aperture on the $F606W$ {\em HST} image and assuming the $F606W$ is identical to the Johnson V.
Then, adopting mass-to-light ratios for V, based on either broadband B-V or V-I colour (Table \ref{t:fg}) and the results of \cite{BelldeJong} in their Figure 9, a stellar mass of $M_* \simeq 3.1 \times 10^9 M_\odot$.
We can also estimate the likely circular velocity via the Tully-Fisher relation. Our V-band luminosity suggests $v_{\rm rot} = 97 ~ km/s$ \citep{Verheijen01b}. All this suggest that the galaxy is a relatively low-mass spiral, comparable to M33 or NGC 4244 and consistent with the dust structure power spectrum observed.

We now estimate the gas mass by first approximating the dust mass, and then assuming a dust-to-gas ratio. We derive the dust mass by assuming the mean optical depth we derived in H09 ($\tau_V = 0.12$) is typical for the entire disk (i.e., the Kron aperture we used above). Implicitly, this assumes that the dark features in front of the background galaxy are not tidal or asymmetric. The dust mass derived this way is a lower limit, as we assume a flat distribution of a finite (truncated) disk, not an exponential one. The latter assumption is predicated on a scale-length of the dust larger than the stellar one, which appears to be valid for smaller disks \citep[][]{MacLachlan11, Holwerda11iau, Holwerda12a}. We then adopt a gas-to-dust ratio of 100 and 500 \citep[typical for lower mass disks in][]{Leroy11}, yielding $M_{\rm gas} \sim 0.5-2.6 \times 10^8 ~ M_\odot$.

Combining the stellar and gas masses, we infer a --rather uncertain-- gas fraction $f_{\rm gas} = M_{\rm gas}/ (M_{\rm gas} + M_*) \sim 0.02-0.08$. This gas fraction for a stellar mass of $M_* \sim 10^9 M_\odot$ is very low compared to typical values \citep{Kannappan04, West10a}, or the general trend of \hi \ gas fraction with stellar mass \citep[][for bigger disks]{Catinella10, Fabello11a}.
However, because of the mass of dust available for shielding within the Kron radius, the bulk of this gas is most likely in the molecular phase. Compared to the trend of molecular gas fraction with stellar mass \citep{Saintonge11a}, the gas fraction is still relatively low, although much less unusual. Assuming an atomic gas envelope of equal mass, in addition to the molecular disk within the Kron radius, the gas fraction would be typical for the measured stellar mass.
Spiral galaxies of similar stellar mass are in the HERACLES survey \citep{Leroy09}, with similar molecular gas mass disks inferred from CO observations, all of which is well contained {\em within} the stellar radius ($R_{\rm 25}$). 

Recent Herschel results on the SED of stacked populations of galaxies point to a low dust temperature ($T_{\rm dust} \sim 15 ~ K$) and low optical/ultraviolet attenuation of the stellar population, together with a high relative dust mass ($M_{\rm dust} / M_*$) from sub-mm emission for blue galaxies with a stellar mass of  $\sim10^9 M_\odot$ \citep{Bourne12a}. The dust geometry we observe in the foreground disk would explain these characteristics: dust is heated less by the stellar disk, because much of the dust is more radially extended that the stars. In addition, the stellar attenuation is not as high for the same reason, yet there is a relatively large amount of cold dust to emit at sub-mm wavelengths. This is reminiscent of a nearby counterpart for this galaxy, NGC 4244 ($v_{\rm rot} = 100 ~ km/s$), which also displays a large dust mass which is distributed in a clumpy medium throughout the disk, right up to the optical radius \citep[$R_{\rm 25}$][]{MacLachlan11,Holwerda11iau, Holwerda12a}.

As discussed above, we suspect this pair is not interacting, although this scenario cannot be ruled out based on the respective redshifts. 
If the pair is physically closely associated, the foreground galaxy's trajectory could take it through the halo of the more massive background galaxy.
This, in turn, could have resulted in the extended dusty disk without the need for a strong tidal interaction. The interaction with the intergalactic medium and gaseous halo of the foreground galaxy would result in a displacement and reprocessing of the ISM disk.
However, if the pair is not interacting, the remaining interpretations are intriguing. What star formation history, in time and galactic radius, would result in such an extended dusty disk? 
Can radial migration \citep[e.g.,][Radburn-Smith et al. {\em in prep}]{Roskar10b} of the stars (inward) or ISM (outward), be the cause of the extend of the dusty disk? 
Or, very speculatively, is the stellar distribution in the core region physically thicker than the outer disk, and the whole galaxy disk is very dusty and opaque \citep[a ``heavy smoker", see][]{Disney89, Phillips91}? More detailed observations of the kinematics of this dusty disk (e.g., with ALMA or MeerKAT) would help settle these questions.

The observed extinction curves can be well fit with a relation described in \cite{CCM} with grayer (lower $R_V$) slopes (Figure \ref{f:ccm:map}). We attribute this predominantly to the sampling of the VIMOS observations. In IFU observations of a closer occulting pair that {\em is} interacting, UGC 3995A/B, we find a similarly gray extinction curve for similar physical sampling ($\sim 9$ kpc, Holwerda et al. {\em in prep}). We note that in that pair, the distribution of $A_V$ values in the attenuation map based on the {\em HST} image display a much higher {\em mean} value than in this foreground spiral, pointing to a fundamental difference in the ISM disk. This difference in $A_V$ distribution between this pair and the interacting UGC 3995A/B, is circumstantial evidence that the foreground galaxy is not distorted by a gravitational interaction with the background galaxy. Alternatively, the still high value of $R_V$ measured over the course VIMOS sampling can be interpreted in tow ways:
either more extended structures with the same attenuation properties exist in the disk (e.g., the spiral structures seen in Figure \ref{f:hst:fov}) or the native $R_V$ values were much higher than the canonical $R_V=3.1$, the signature of recently produced or reprocessed dust, circumstantial evidence of recent interaction.

Future work on occulting galaxy pairs will benefit greatly from higher spatial sampling of the next generation of IFU, e.g., the WEAVE instrument on the WHT or MUSE on VLT, both assisted with AO systems. To fully answer the question whether dust disks such as the one observed in the foreground disk is typical, a statistical sample of occulting galaxy pairs observed with an IFU instrument will be needed.

\section*{Acknowledgements}

The authors would like to thank the anonymous referee for his or her comments and suggestions for the improvement of the manuscript.
The lead author thanks the European Space Agency for the support of the Research Fellowship program.
We thank the ESO staff for their support of the service observation program and the anonymous referee for his or her very helpful comments and suggestions.
Based on VIMOS observations carried out using the Very Large Telescope at the ESO Paranal Observatory under Programme ID: 384.B-0059A, as well as observations made with the NASA/ESA Hubble Space Telescope, which is a collaboration between the Space Telescope Science Institute (STScI/NASA), the Space Telescope European Coordinating Facility (ST-ECF/ESA), and the Canadian Astronomy Data Centre (CADC/NRC/CSA). The Hubble data presented in this paper were obtained from the Multimission Archive at the Space Telescope Science Institute (MAST). STScI is operated by the Association of Universities for Research in Astronomy, Inc., under NASA contract NAS5-26555. Support for MAST for non-HST data is provided by the NASA Office of Space Science via grant NAG5-7584, and by other grants and contracts.
This research has made use of the NASA/IPAC Extragalactic Database (NED) which is operated by the Jet Propulsion Laboratory, California Institute of Technology, under contract with the National Aeronautics and Space Administration and NASA's Astrophysics Data System (ADS).

This paper made use of the {\sc matplotlib} \citep{matplotlib} plotting package under the {\em Python} environment.

\clearpage
\newpage

\appendix

\begin{table*}
{\tiny
\section{Spectral smoothing and the CCM fit}
\caption{The normalization ($A_V$) and slope ($R_V$) of the CCM fit to the attenuation curves in Aperture II, inferred from the raw and spectrally smoothed spectra (5,7,9,11 and 21 pixels respectively).}
\begin{center}
\begin{tabular}{l l l l l l l l l l l l l l l l l l l l}
Fiber 	& Raw & & & 13 \AA & & & 18.2 \AA & & & 23.4 \AA & & & 28.6 \AA & & & 54.6 \AA & \\
No.		& $A_V$ & $R_V$ & $\chi^2$ & $A_V$ & $R_V$ & $\chi^2$ &$A_V$ & $R_V$ & $\chi^2$ &$A_V$ & $R_V$ & $\chi^2$ &$A_V$ & $R_V$ & $\chi^2$ &$A_V$ & $R_V$ & $\chi^2$ \\
\hline
\hline

1	& 0.59 	 & 2.52 	 & 2.07 & 0.59 	 & 2.53 	 & 1.19 & 0.59 	 & 2.53 	 & 0.91 & 0.59 	 & 2.53 	 & 0.72 & 0.59 	 & 2.53 	 & 0.59 & 0.59 	 & 2.53 	 & 0.34 \\ 
2	& 0.60 	 & 1.71 	 & 6.16 & 0.60 	 & 1.71 	 & 3.36 & 0.60 	 & 1.71 	 & 2.56 & 0.60 	 & 1.71 	 & 2.05 & 0.60 	 & 1.71 	 & 1.69 & 0.60 	 & 1.70 	 & 0.85 \\ 
3	& 0.67 	 & 1.24 	 & 24.48 & 0.66 	 & 1.24 	 & 14.42 & 0.66 	 & 1.24 	 & 11.17 & 0.66 	 & 1.23 	 & 9.19 & 0.66 	 & 1.23 	 & 7.87 & 0.66 	 & 1.21 	 & 4.92 \\ 
4	& 0.70 	 & 2.96 	 & 32.08 & 0.69 	 & 3.13 	 & 14.92 & 0.69 	 & 3.18 	 & 10.34 & 0.69 	 & 3.21 	 & 7.83 & 0.69 	 & 3.23 	 & 6.26 & 0.69 	 & 3.28 	 & 3.00 \\ 
5	& 0.52 	 & 1.43 	 & 37.72 & 0.51 	 & 1.48 	 & 19.16 & 0.51 	 & 1.49 	 & 13.51 & 0.51 	 & 1.50 	 & 10.09 & 0.51 	 & 1.50 	 & 7.88 & 0.51 	 & 1.50 	 & 4.07 \\ 
6	& 0.84 	 & 2.40 	 & 3.54 & 0.84 	 & 2.41 	 & 2.19 & 0.84 	 & 2.41 	 & 1.83 & 0.84 	 & 2.41 	 & 1.62 & 0.84 	 & 2.41 	 & 1.49 & 0.84 	 & 2.41 	 & 1.20 \\ 
7	& 0.63 	 & 5.26 	 & 11.34 & 0.63 	 & 5.46 	 & 6.65 & 0.63 	 & 5.50 	 & 5.40 & 0.63 	 & 5.50 	 & 4.71 & 0.63 	 & 5.49 	 & 4.27 & 0.63 	 & 5.46 	 & 3.17 \\ 
8	& 0.62 	 & 0.84 	 & 19.21 & 0.61 	 & 0.84 	 & 9.98 & 0.61 	 & 0.84 	 & 7.55 & 0.61 	 & 0.84 	 & 6.18 & 0.61 	 & 0.84 	 & 5.33 & 0.61 	 & 0.85 	 & 3.50 \\ 
9	& 0.32 	 & 13.68 	 & 23.95 & 0.32 	 & 13.49 	 & 12.92 & 0.32 	 & 13.35 	 & 9.66 & 0.32 	 & 13.18 	 & 7.75 & 0.32 	 & 13.05 	 & 6.52 & 0.32 	 & 13.33 	 & 3.87 \\ 
10	& 0.23 	 & 2.84 	 & 31.16 & 0.23 	 & 2.79 	 & 16.18 & 0.23 	 & 2.75 	 & 11.99 & 0.23 	 & 2.71 	 & 9.77 & 0.23 	 & 2.68 	 & 8.37 & 0.23 	 & 2.61 	 & 5.02 \\ 
11	& 0.45 	 & 1.18 	 & 11.06 & 0.44 	 & 1.18 	 & 8.25 & 0.44 	 & 1.19 	 & 7.30 & 0.45 	 & 1.19 	 & 6.67 & 0.45 	 & 1.19 	 & 6.22 & 0.45 	 & 1.20 	 & 4.93 \\ 
12	& 0.91 	 & 2.71 	 & 14.84 & 0.91 	 & 2.76 	 & 7.82 & 0.91 	 & 2.78 	 & 5.76 & 0.91 	 & 2.80 	 & 4.45 & 0.90 	 & 2.81 	 & 3.54 & 0.90 	 & 2.82 	 & 1.47 \\ 
13	& 0.35 	 & 1.04 	 & 28.77 & 0.35 	 & 1.04 	 & 14.82 & 0.35 	 & 1.04 	 & 10.99 & 0.35 	 & 1.04 	 & 8.94 & 0.35 	 & 1.04 	 & 7.69 & 0.35 	 & 1.03 	 & 4.93 \\ 
14	& 0.27 	 & -0.98 	 & 23.60 & 0.27 	 & -0.98 	 & 13.32 & 0.27 	 & -1.00 	 & 10.44 & 0.27 	 & -1.01 	 & 8.86 & 0.27 	 & -1.02 	 & 7.84 & 0.27 	 & -1.05 	 & 5.60 \\ 
15	& -0.55 	 & 140.69 	 & 28.56 & -0.54 	 & -125.89 	 & 14.87 & -0.54 	 & -79.18 	 & 11.09 & -0.54 	 & -65.04 	 & 8.96 & -0.54 	 & -58.40 	 & 7.55 & -0.54 	 & -52.03 	 & 4.54 \\ 

\hline
\end{tabular}
\end{center}
\label{t:ccm}
}
\end{table*}%

\begin{figure*}
\begin{center}
\includegraphics[width=\textwidth]{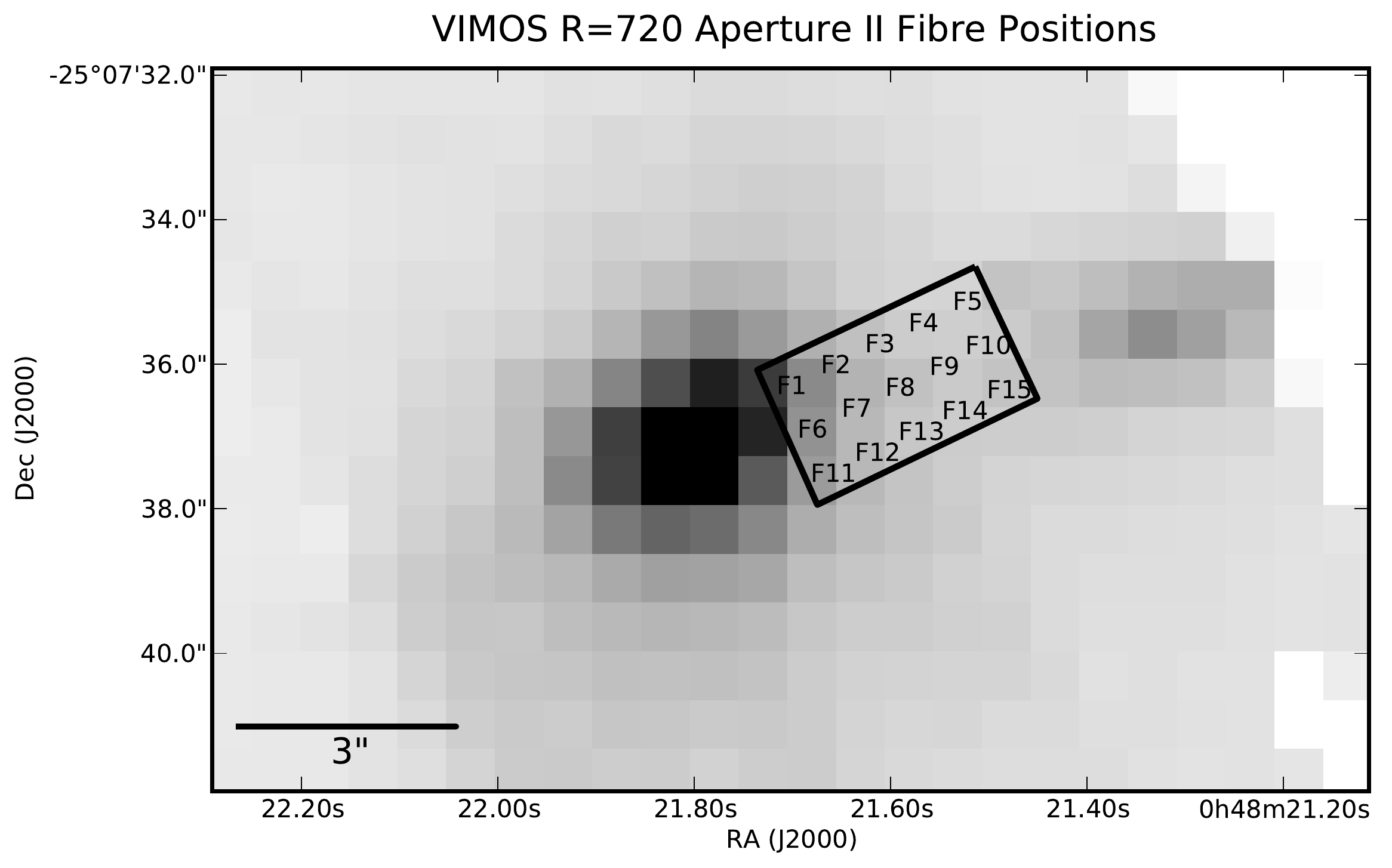}

\caption{The position of the fibres in Aperture II reported in Table \ref{t:ccm} and Figure \ref{f:aper2:fib}. Grayscale was adjusted so clarify fibre labels. }
\label{f:aper2:fibrepos}
\end{center}
\end{figure*}

\begin{figure*}
\begin{center}
\includegraphics[width=0.45\textwidth]{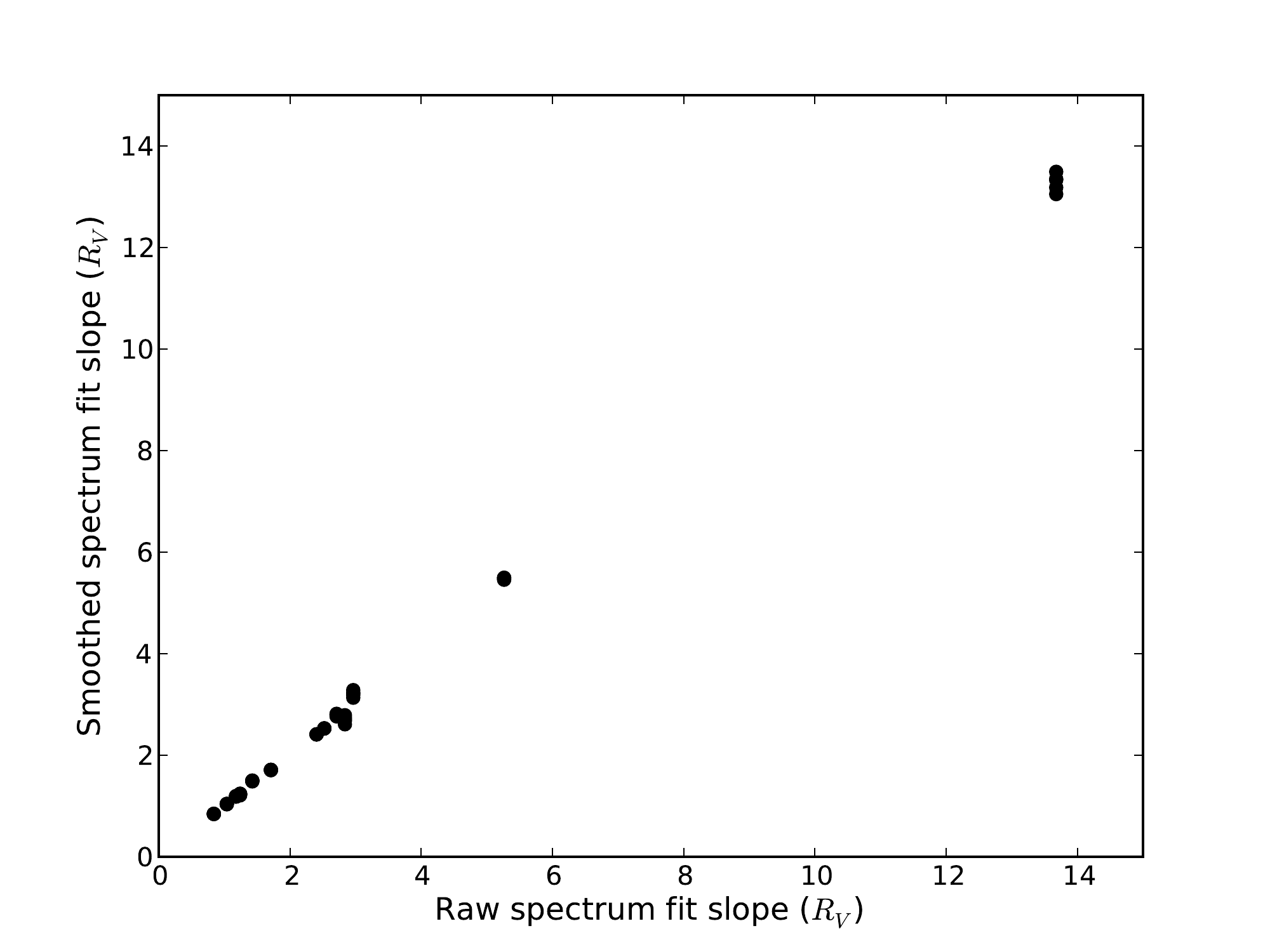}
\includegraphics[width=0.45\textwidth]{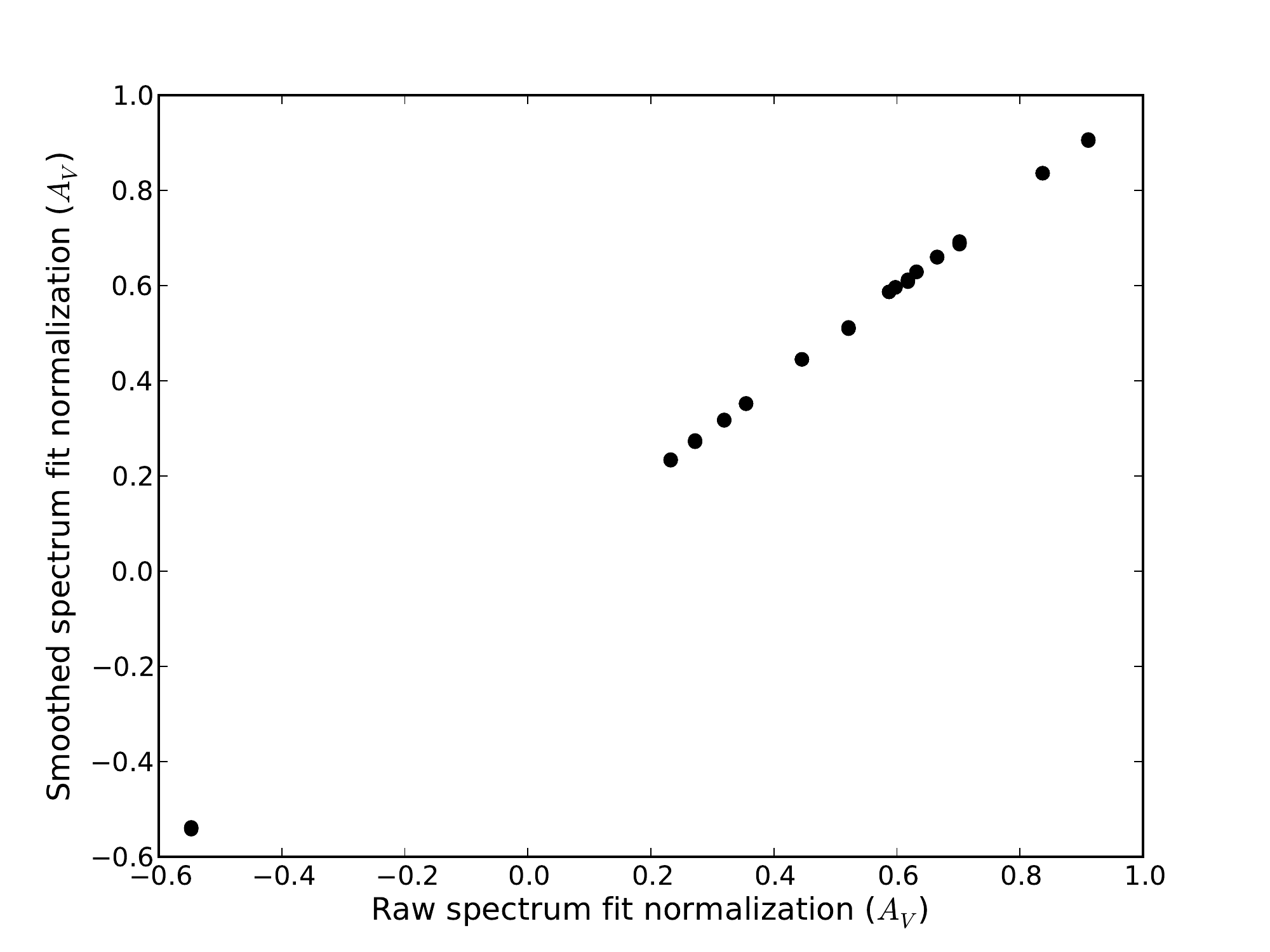}
\caption{Comparison between the raw and spectrally smoothed CCM fits, the slope ($R_V$) and normalization ($A_V$). }
\label{f:ccm}
\end{center}
\end{figure*}

\begin{figure*}
\section{The effects of spectral smoothing}
\begin{center}
\includegraphics[width=0.45\textwidth]{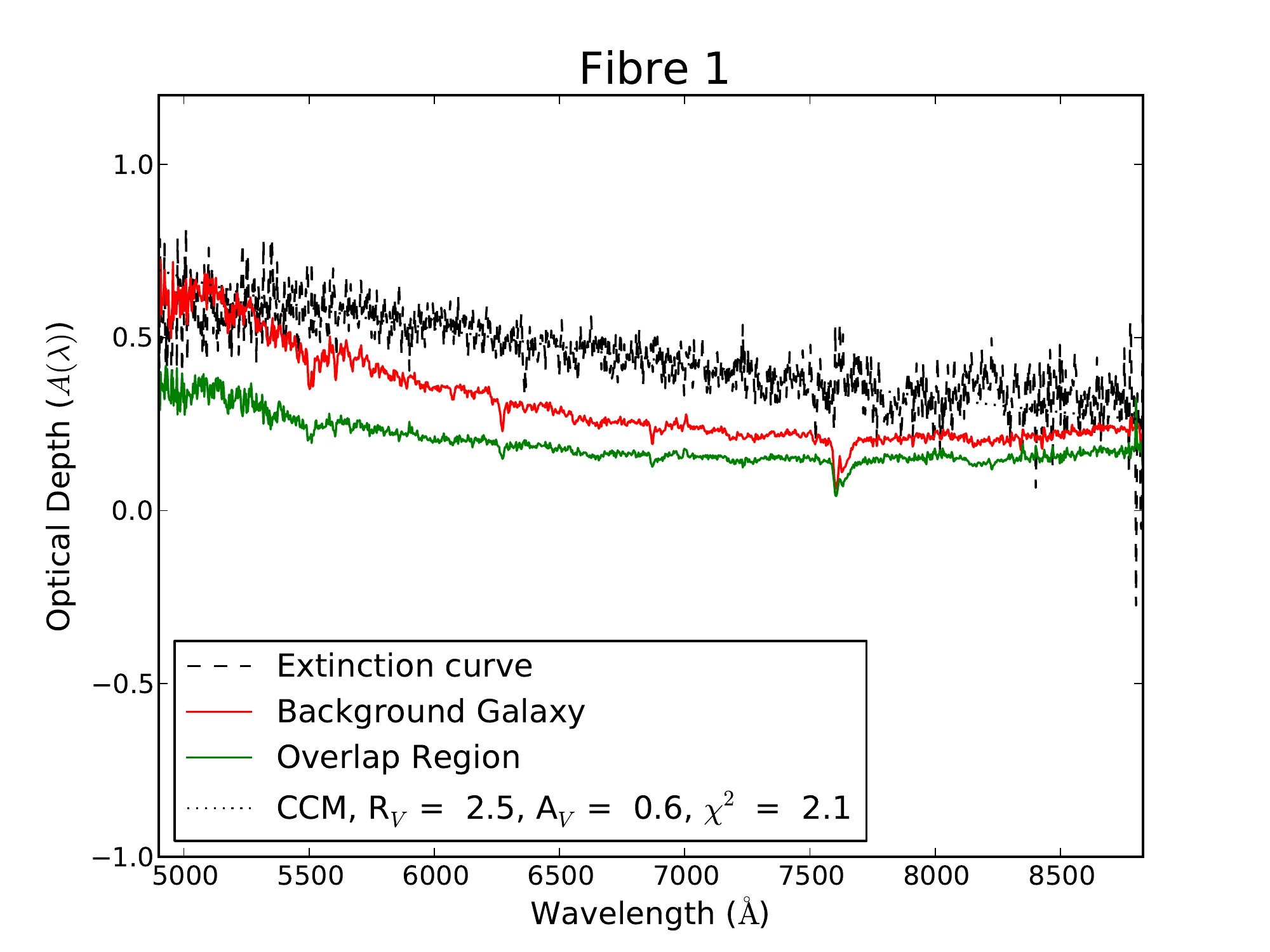}
\includegraphics[width=0.45\textwidth]{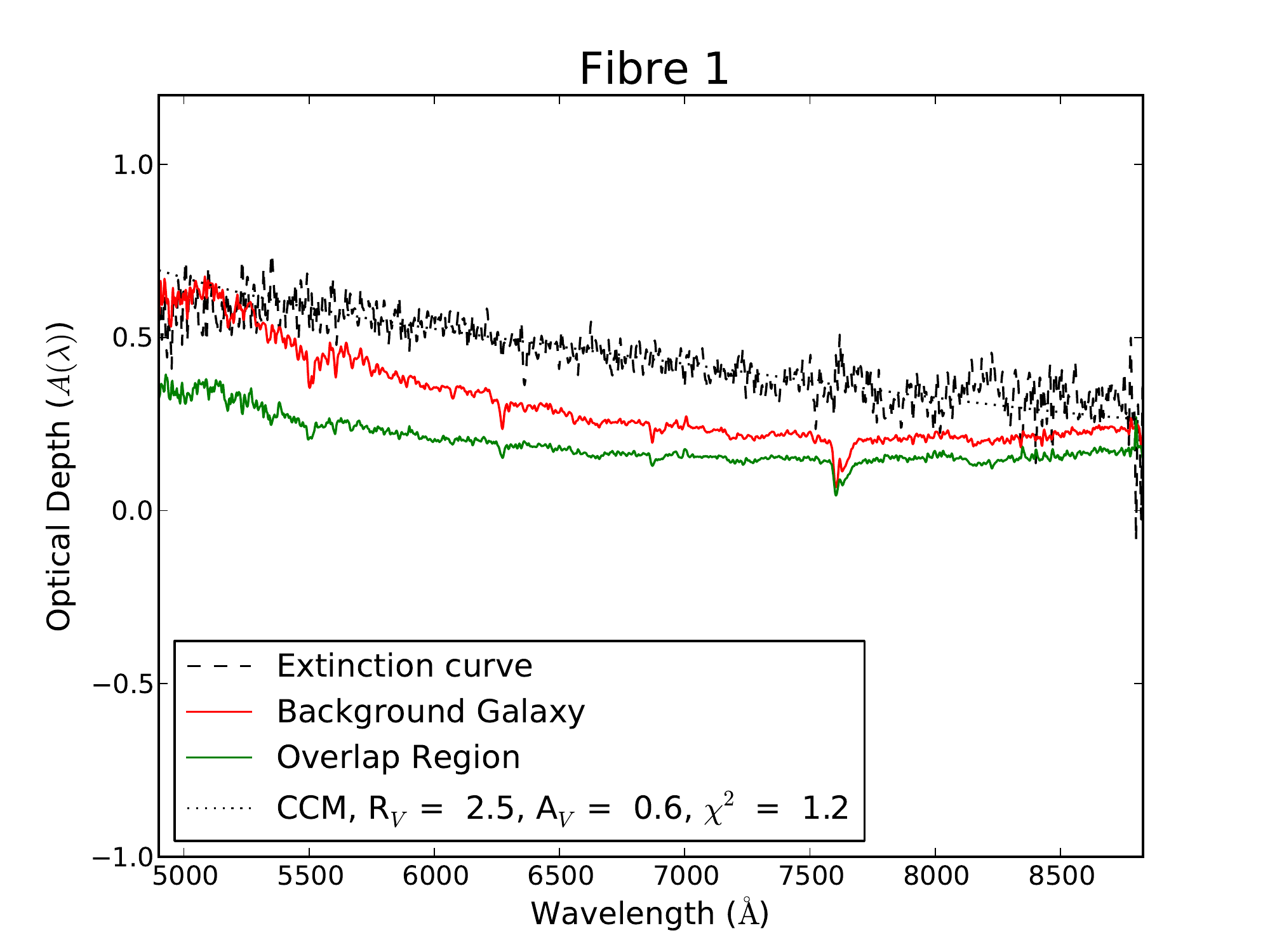}
\includegraphics[width=0.45\textwidth]{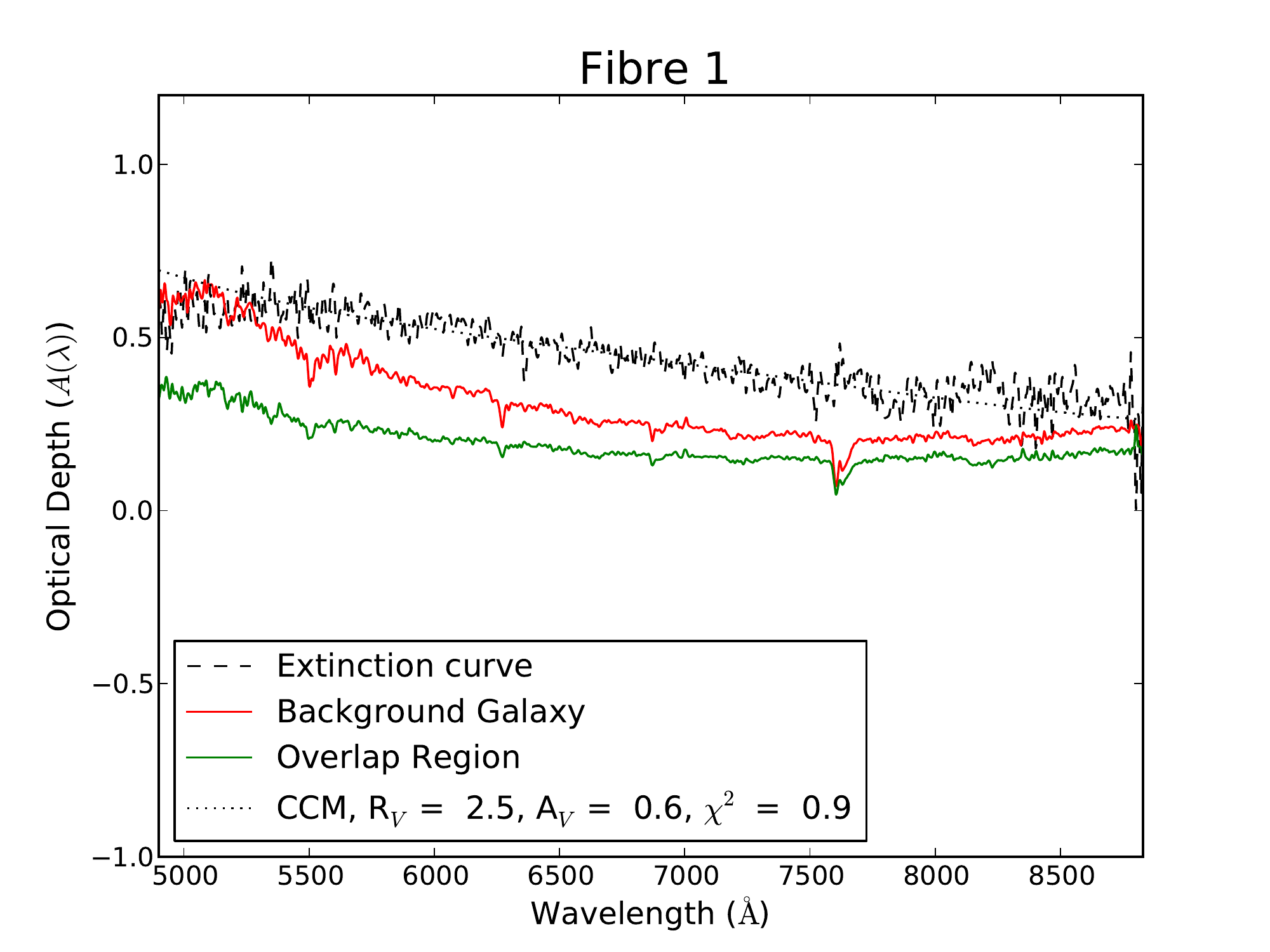}
\includegraphics[width=0.45\textwidth]{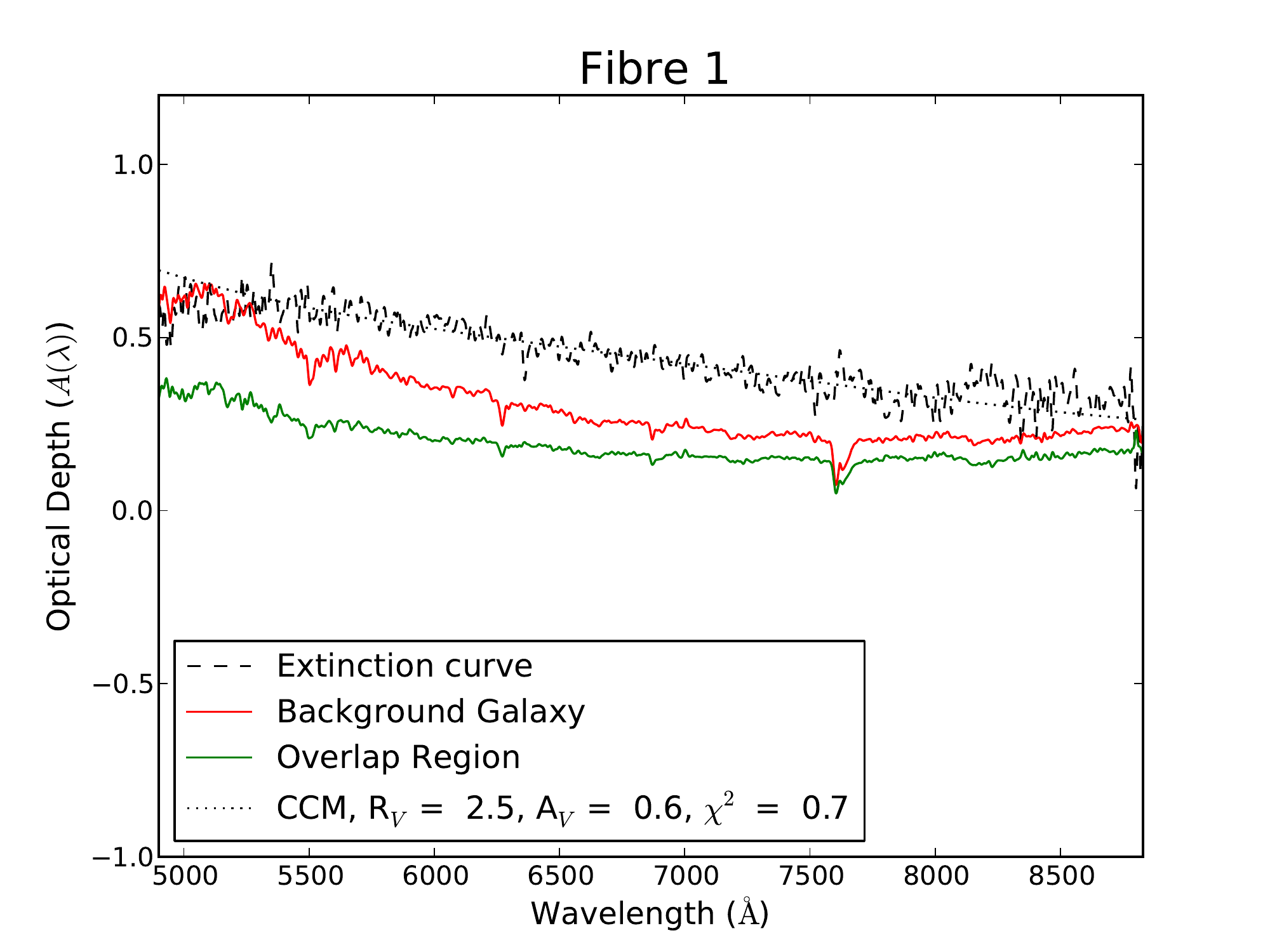}
\includegraphics[width=0.45\textwidth]{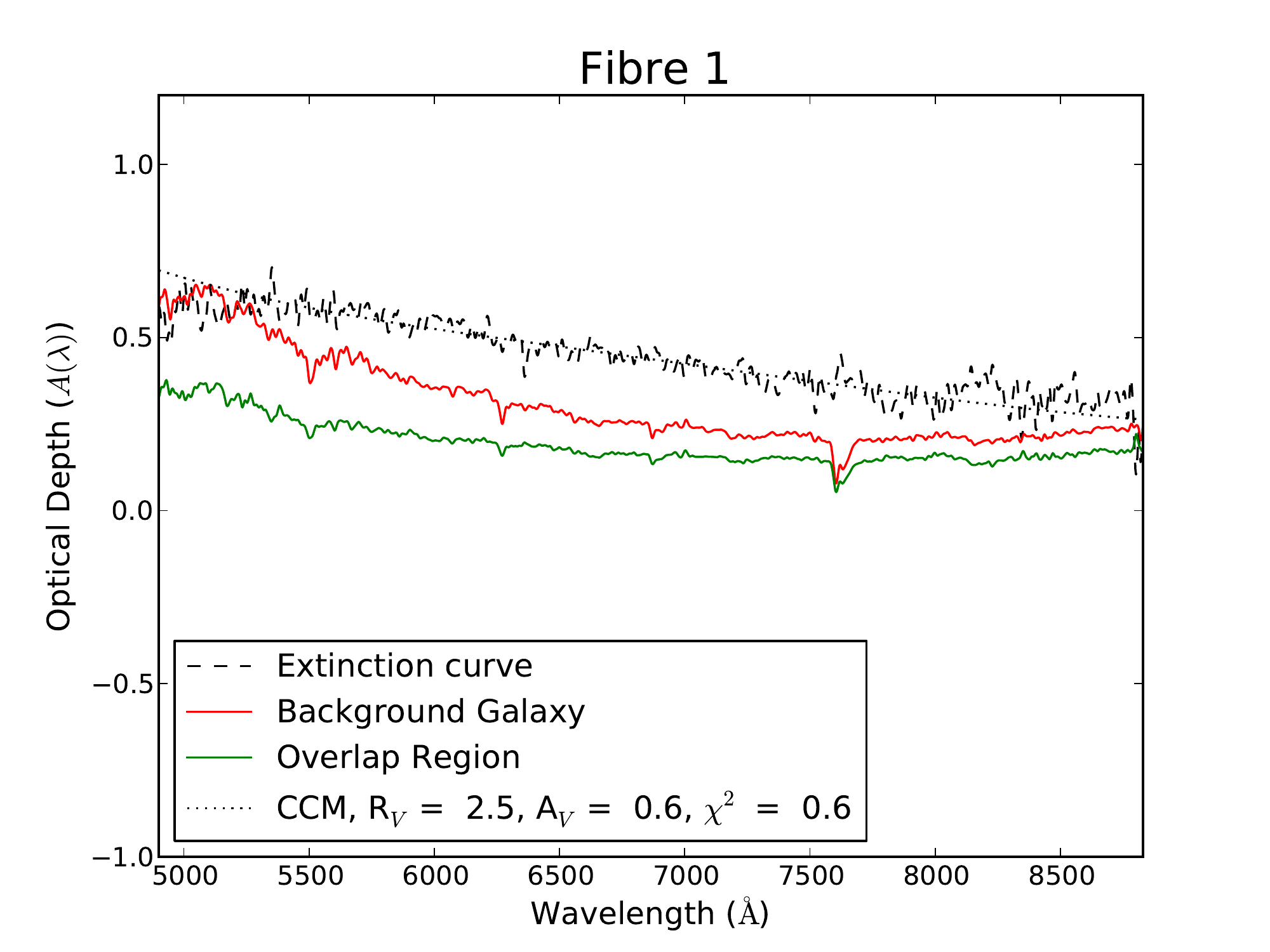}
\includegraphics[width=0.45\textwidth]{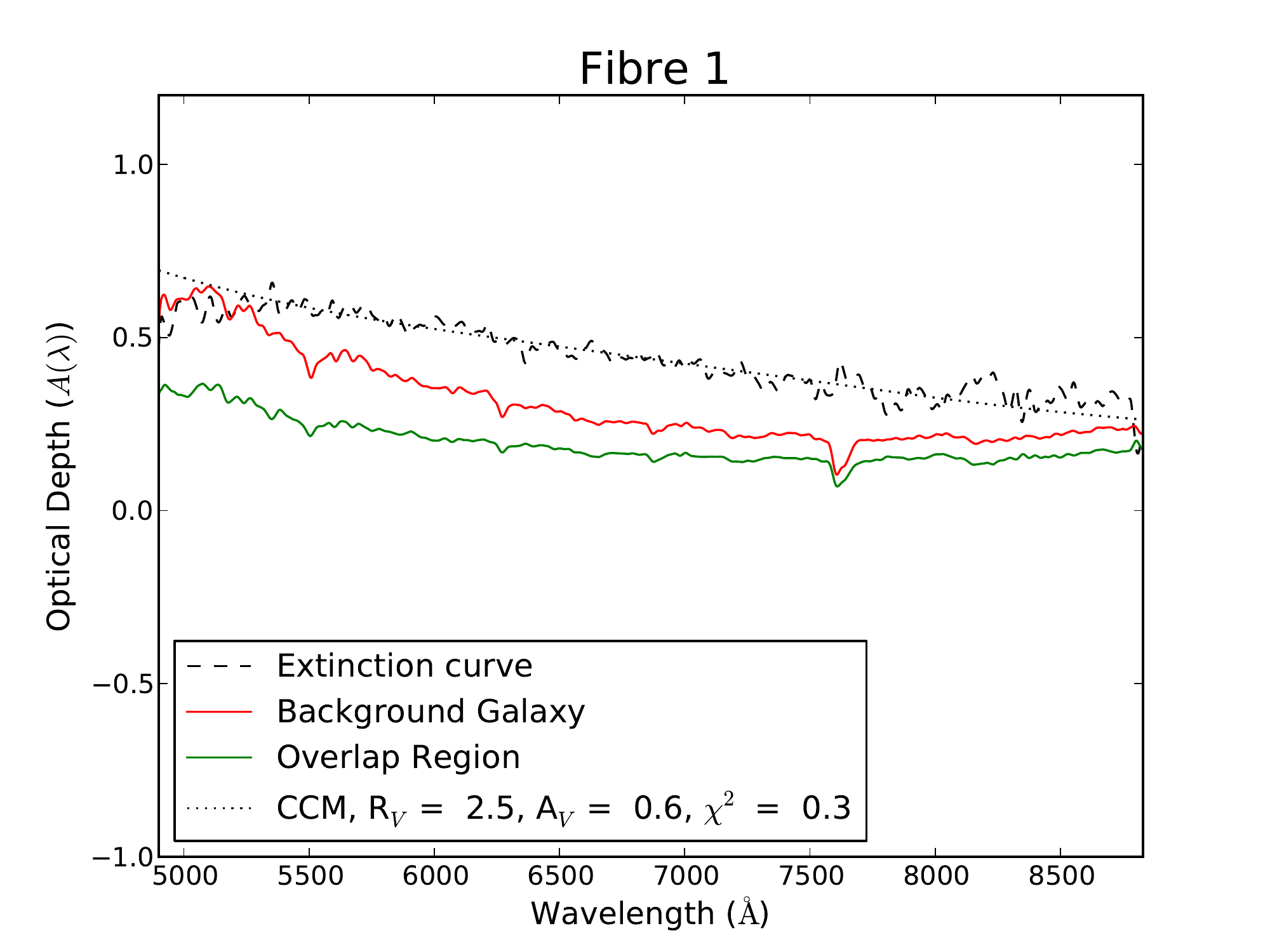}

\caption{An example fibre from Aperture II, smoothed with different Hanning windows (13, 18.2, 23.4, 28.6, and 54.6 \AA) to illustrate the effect of smoothing the spectra to different resolutions. Our CCM fit from the raw fibres are robust in this Aperture and there is no need to smooth in the spectral direction to improve fit.}
\label{f:aper2:smoothing}
\end{center}
\end{figure*}

\begin{figure*}
\begin{center}
\includegraphics[width=0.32\textwidth]{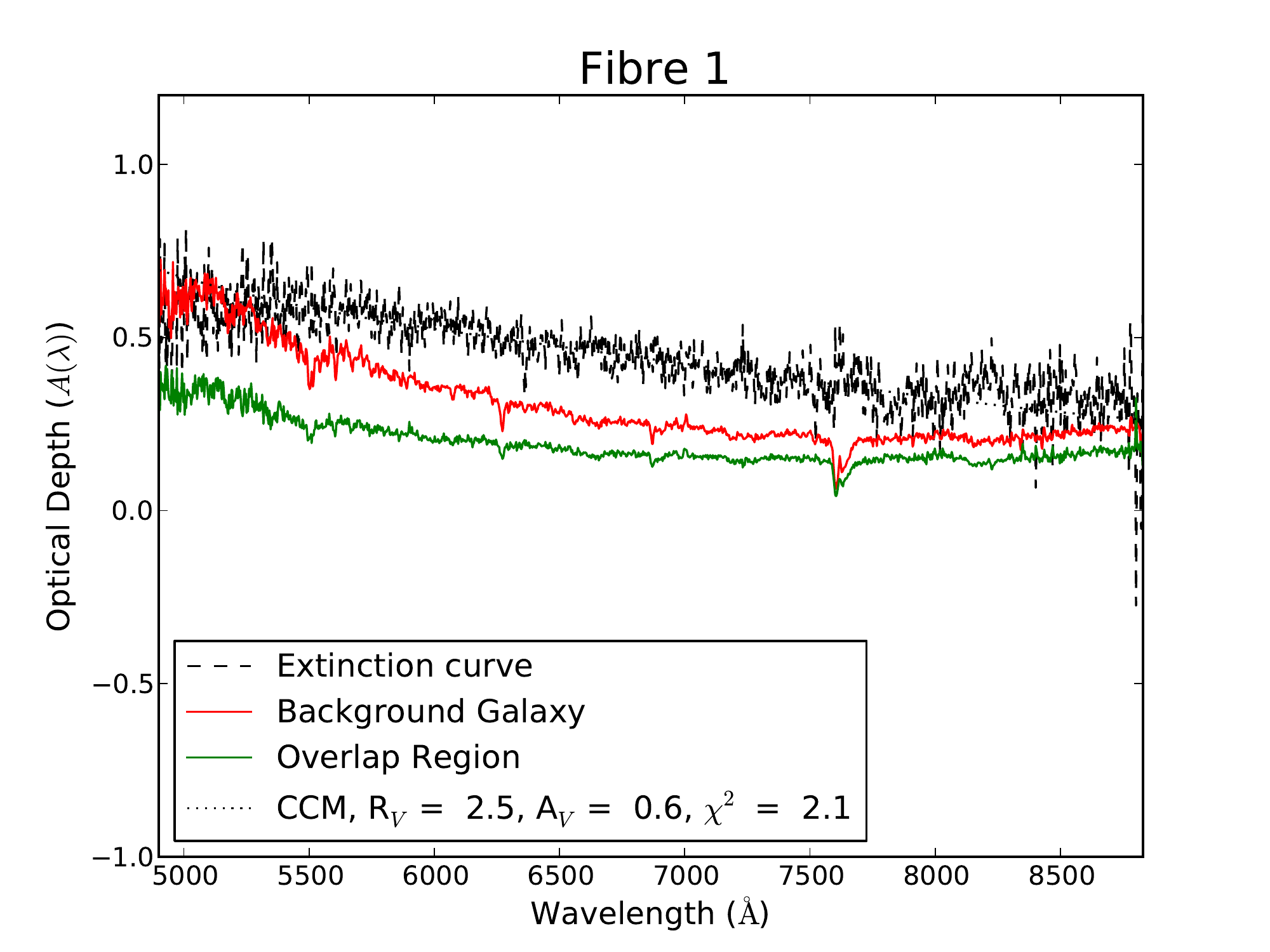}
\includegraphics[width=0.32\textwidth]{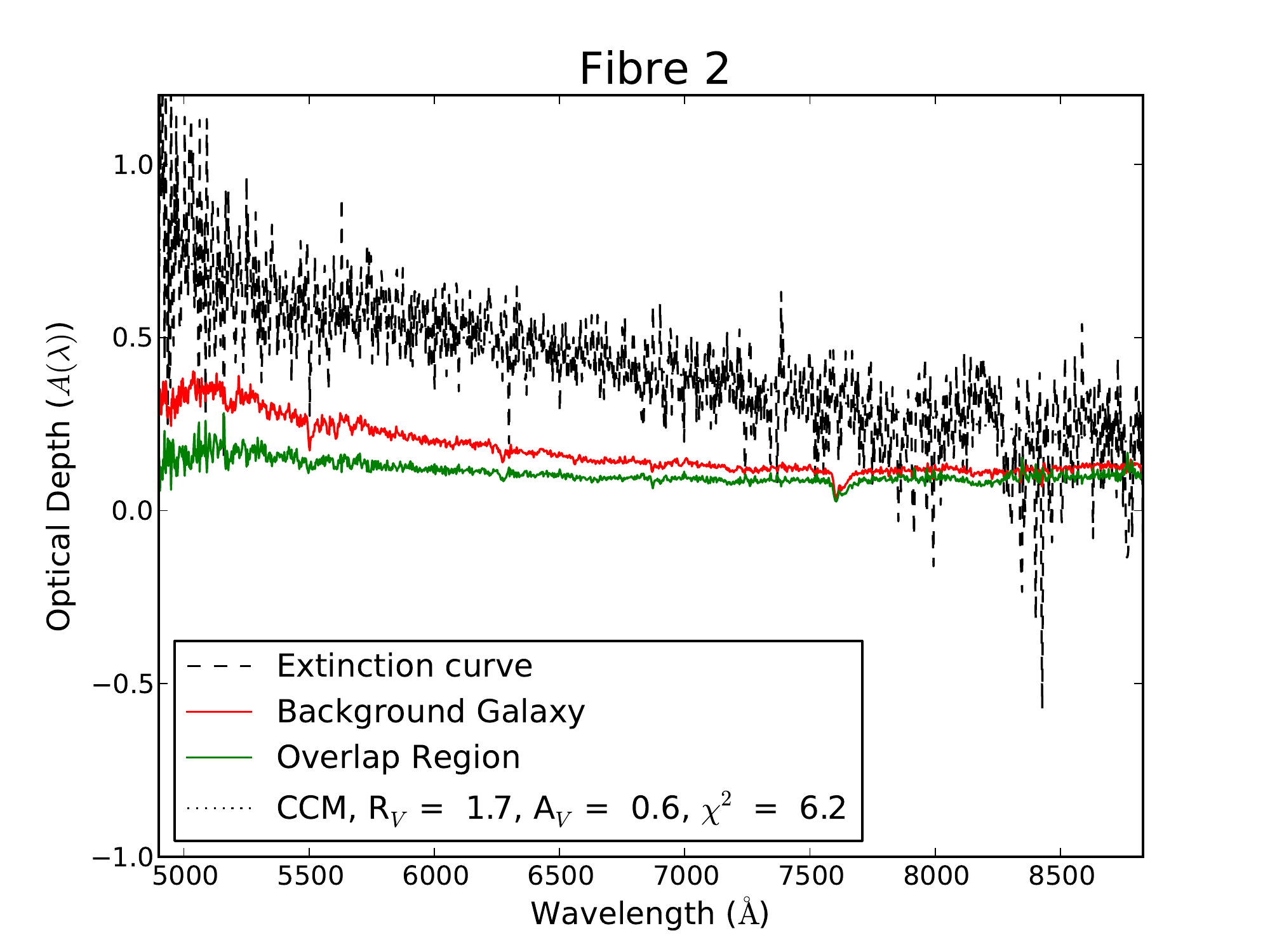}
\includegraphics[width=0.32\textwidth]{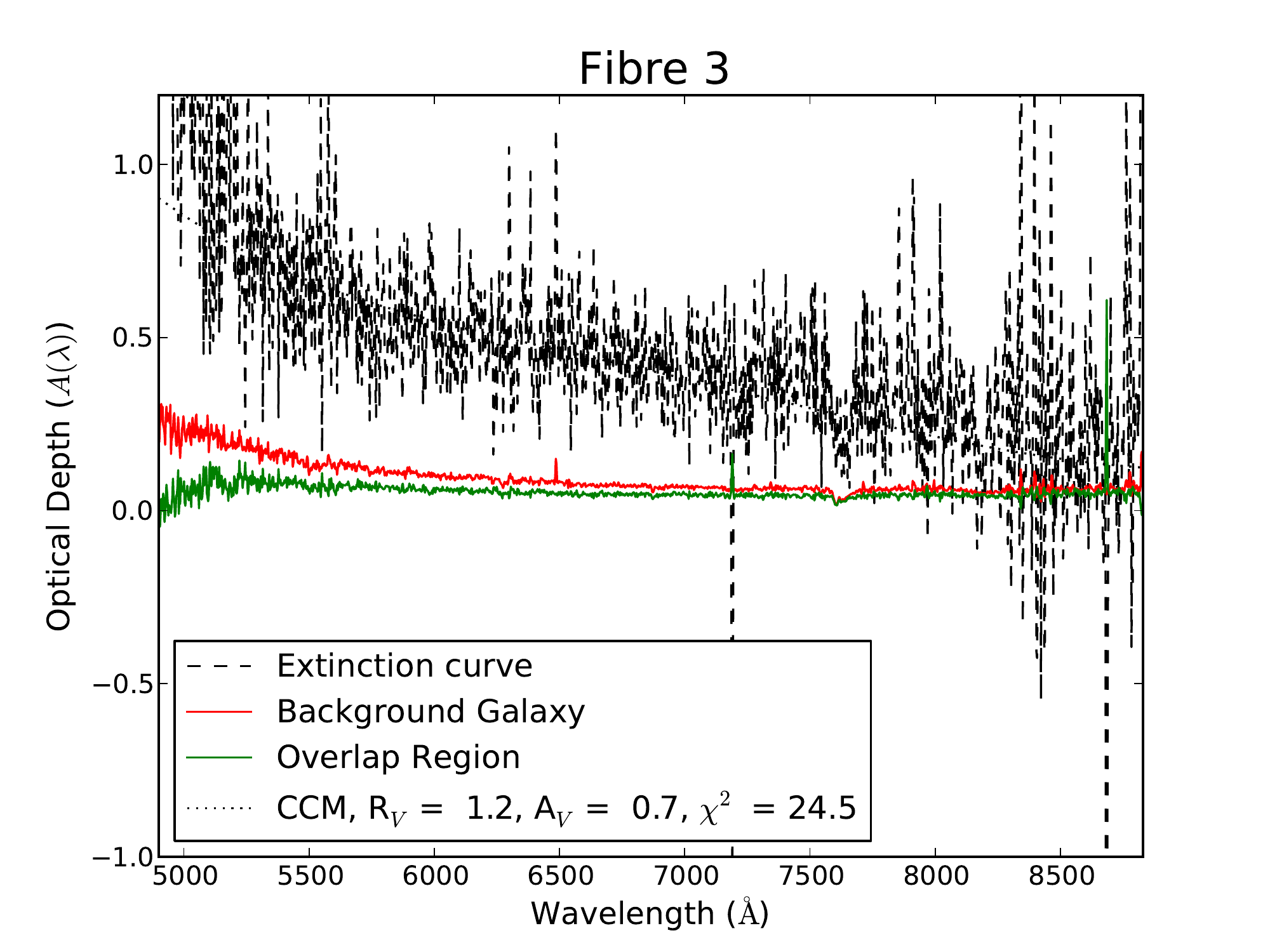}
\includegraphics[width=0.32\textwidth]{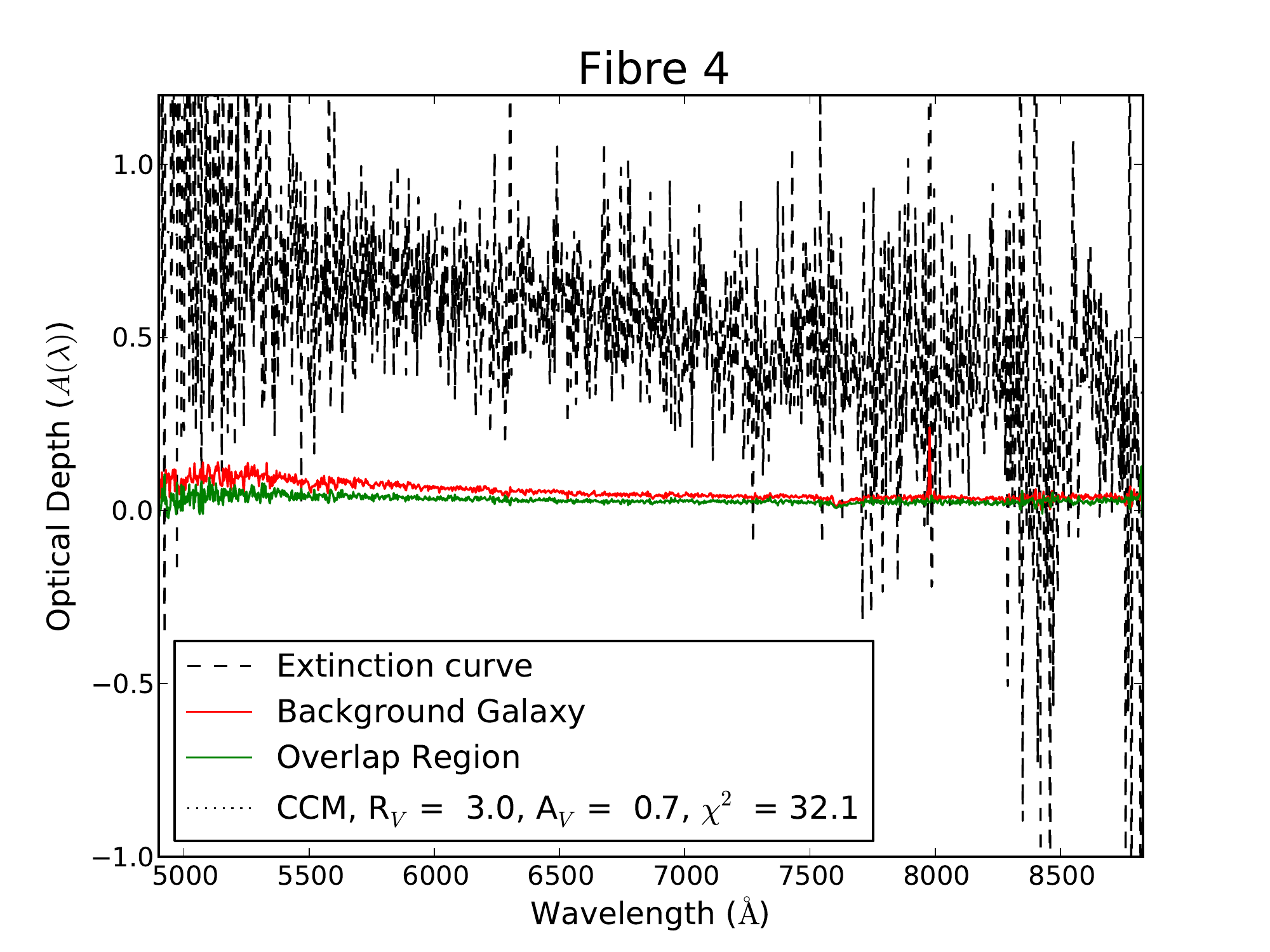}
\includegraphics[width=0.32\textwidth]{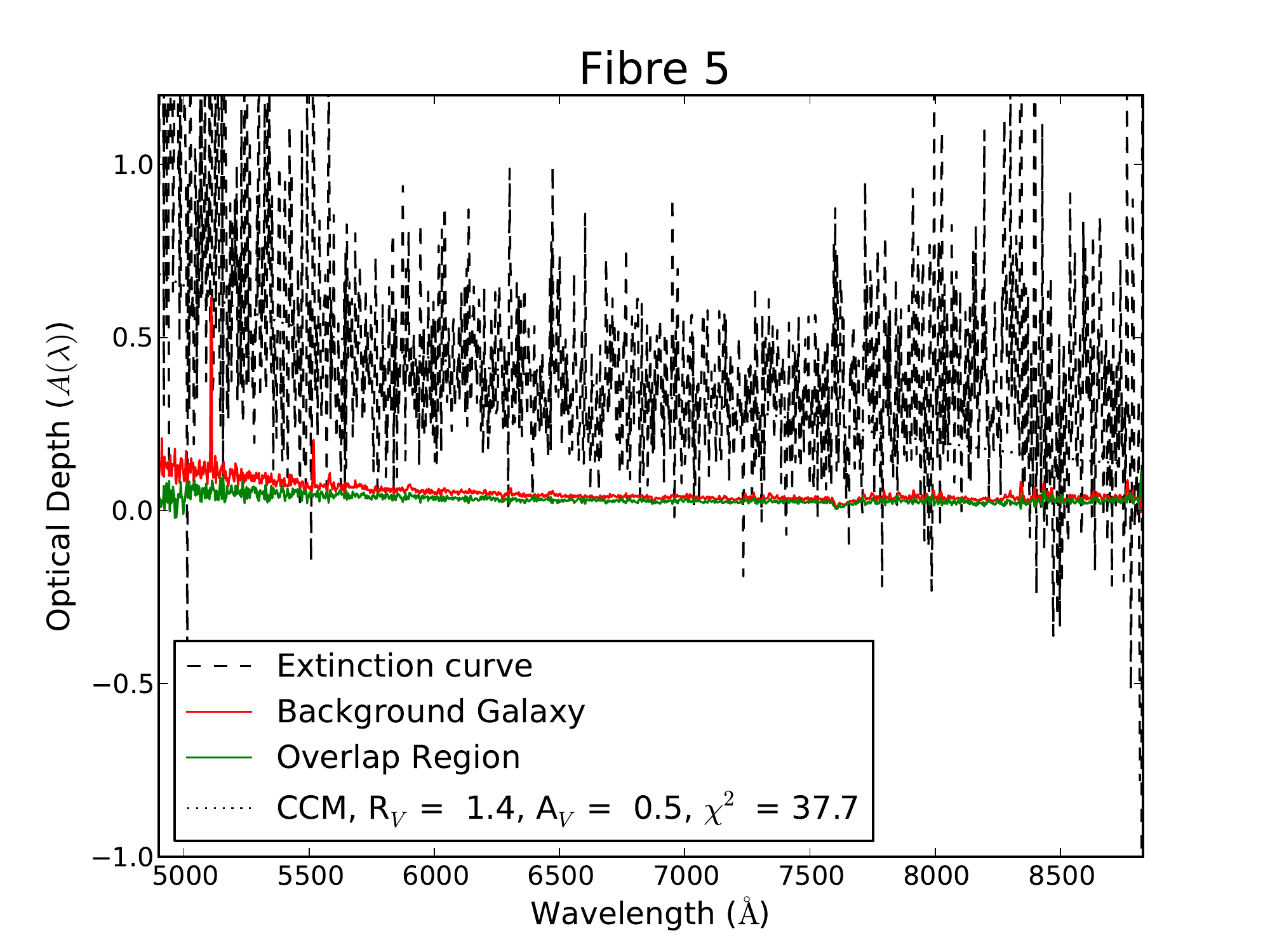}
\includegraphics[width=0.32\textwidth]{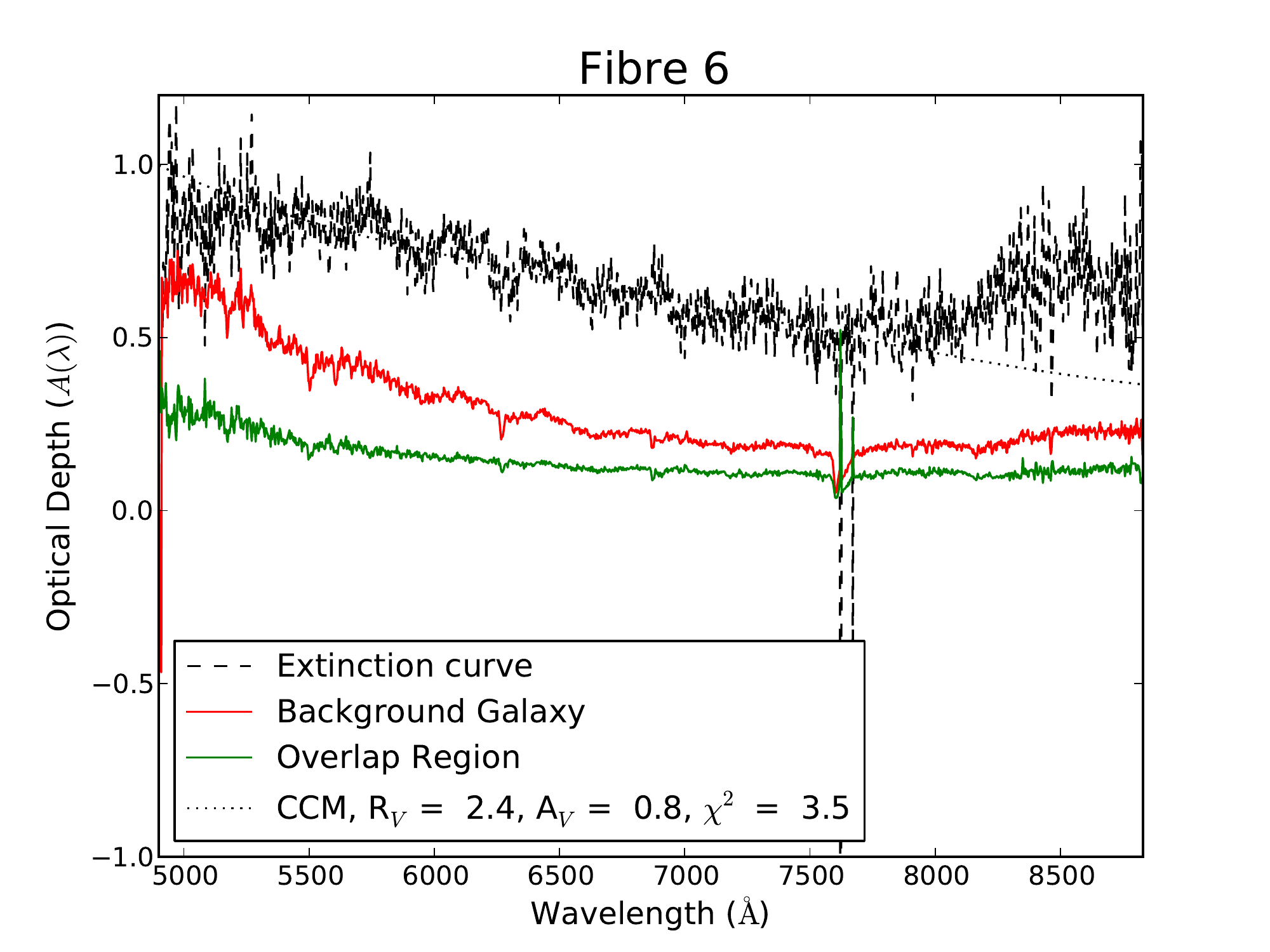}
\includegraphics[width=0.32\textwidth]{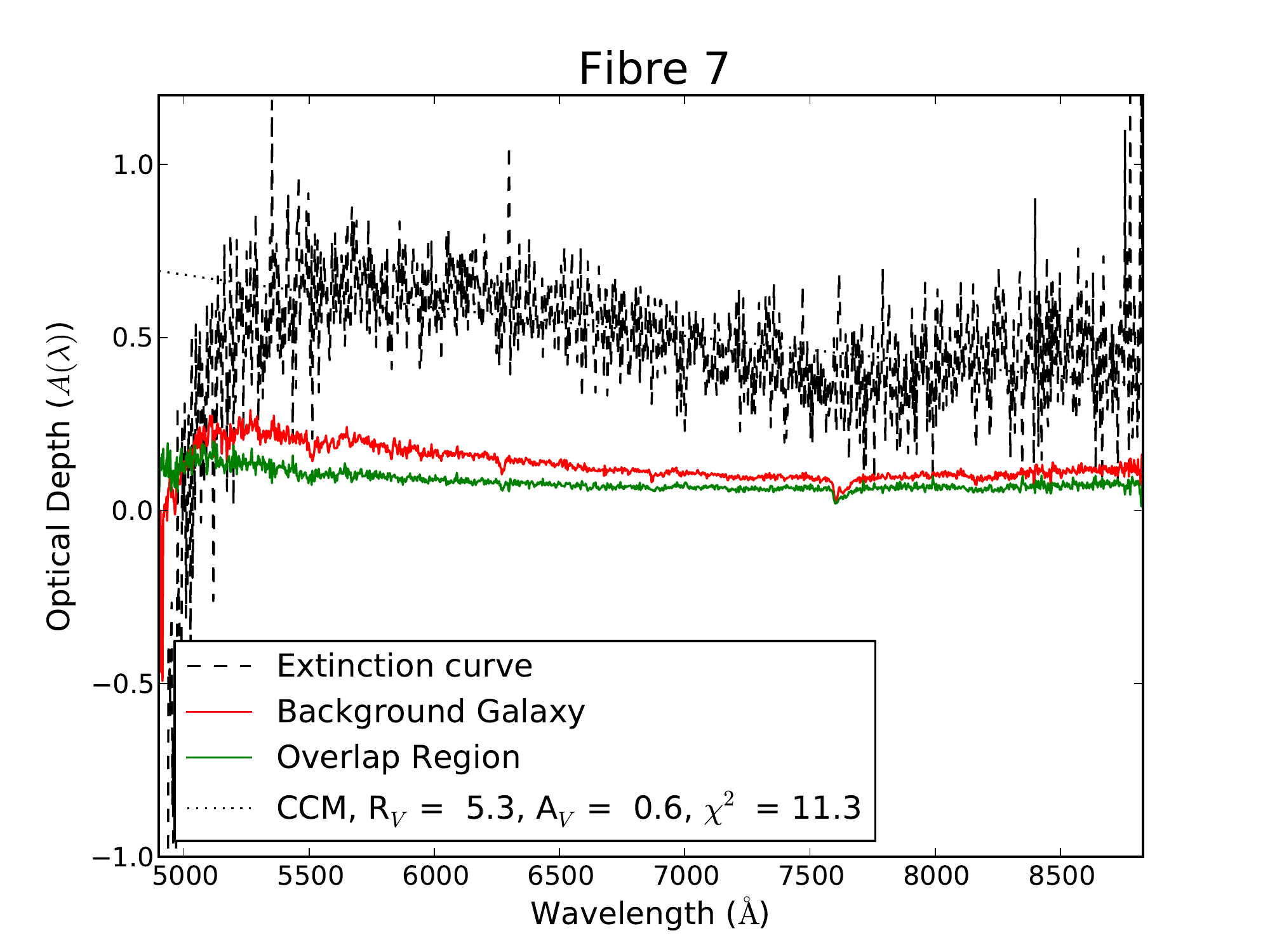}
\includegraphics[width=0.32\textwidth]{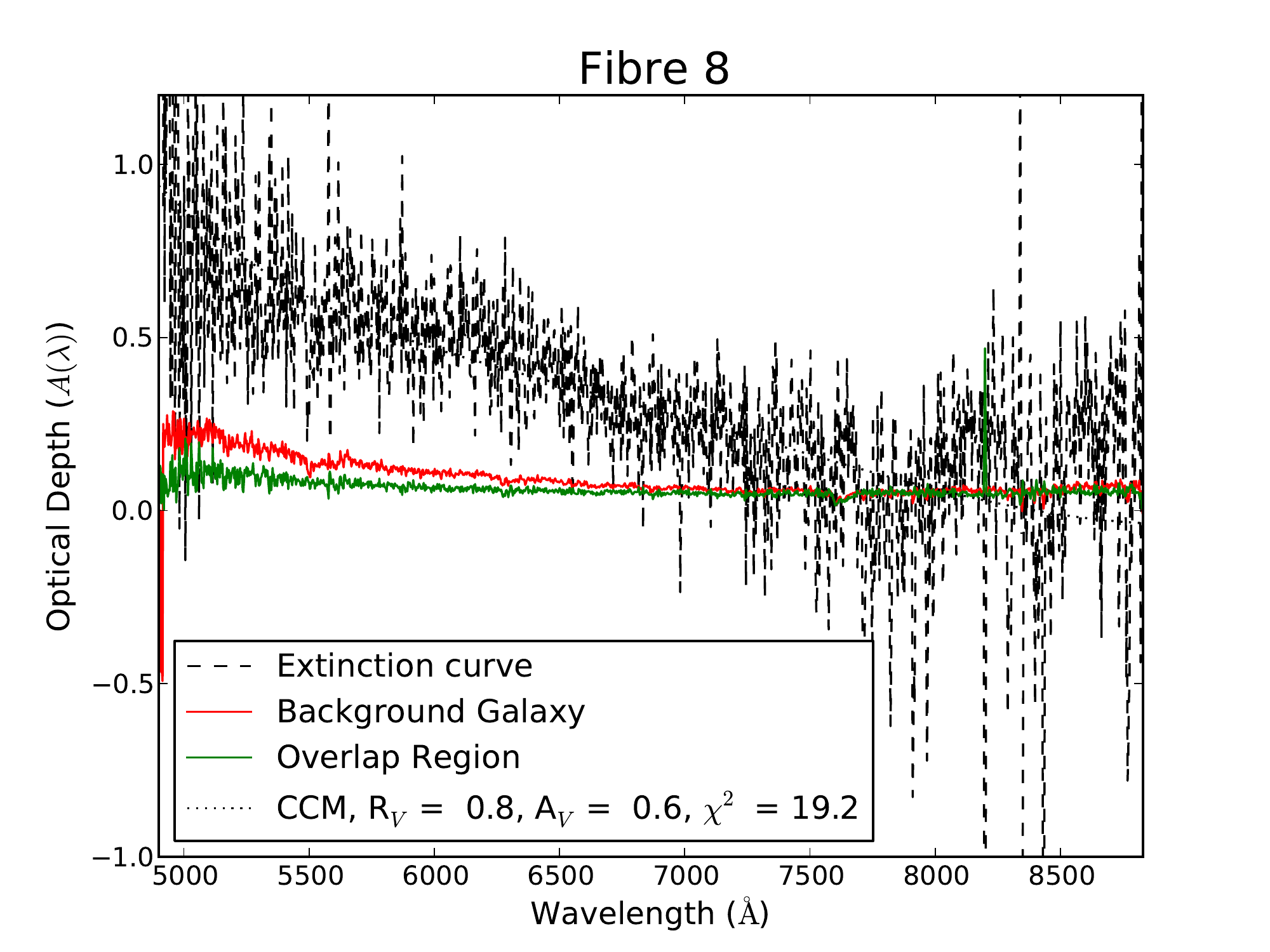}
\includegraphics[width=0.32\textwidth]{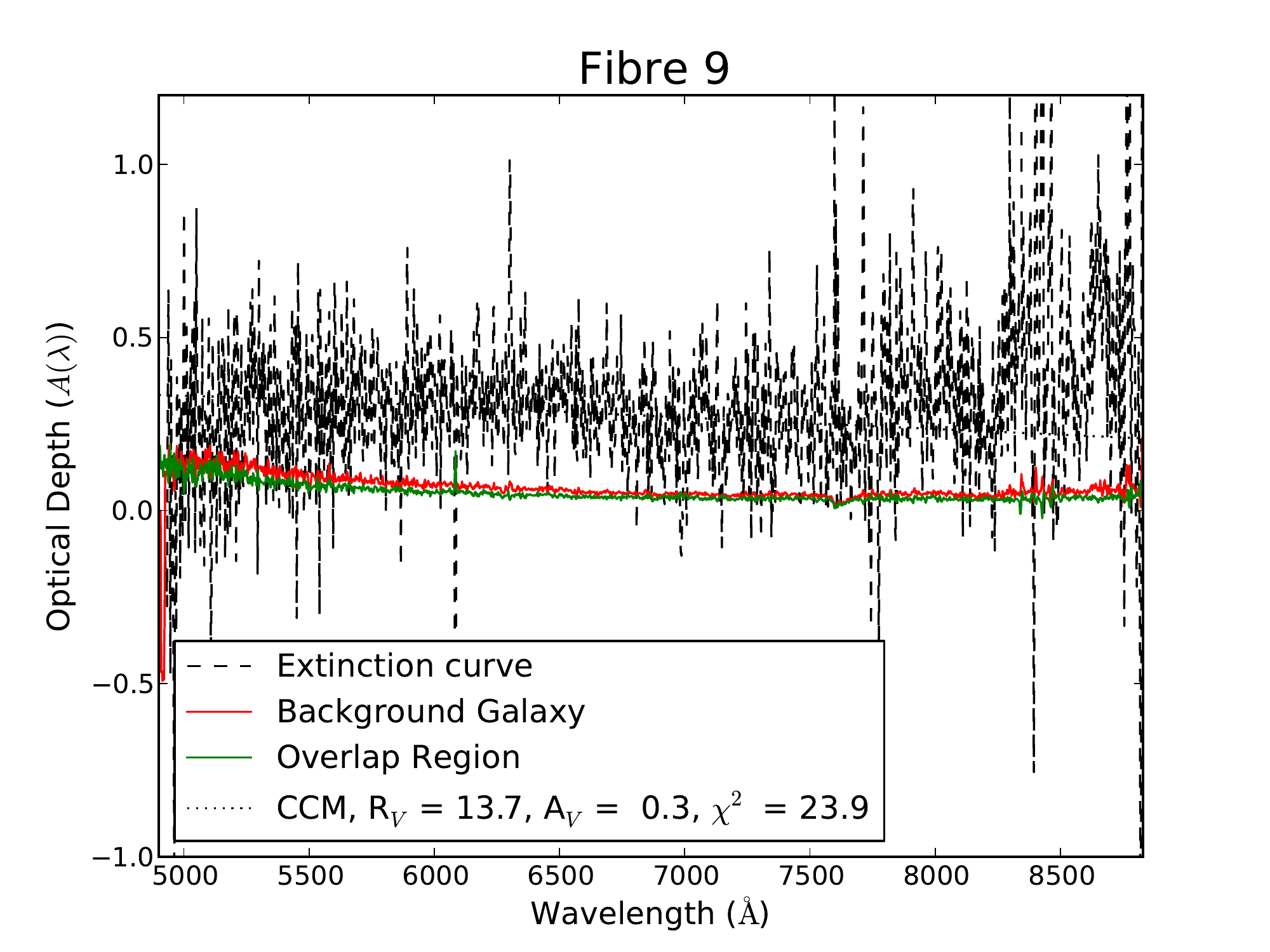}
\includegraphics[width=0.32\textwidth]{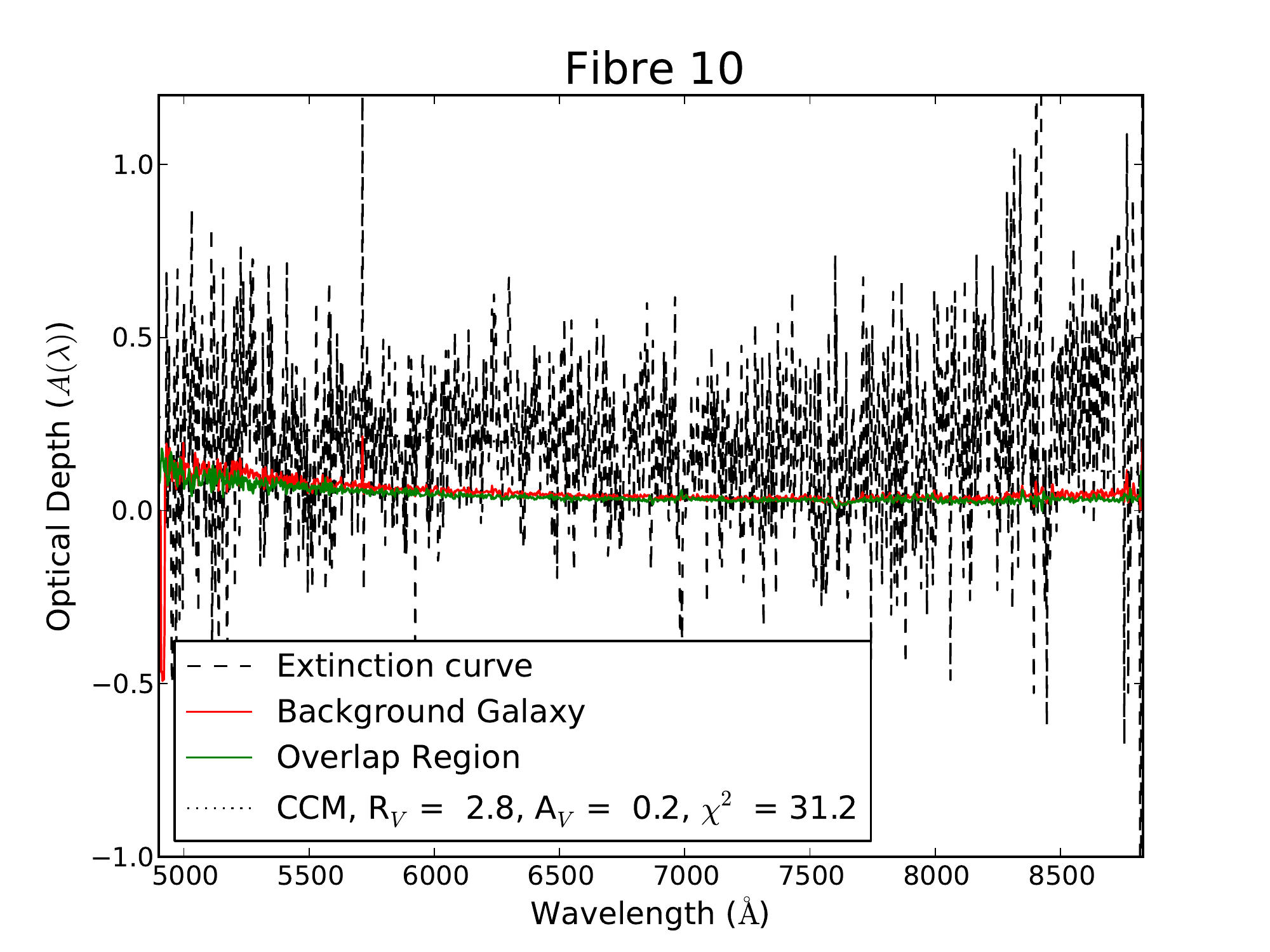}
\includegraphics[width=0.32\textwidth]{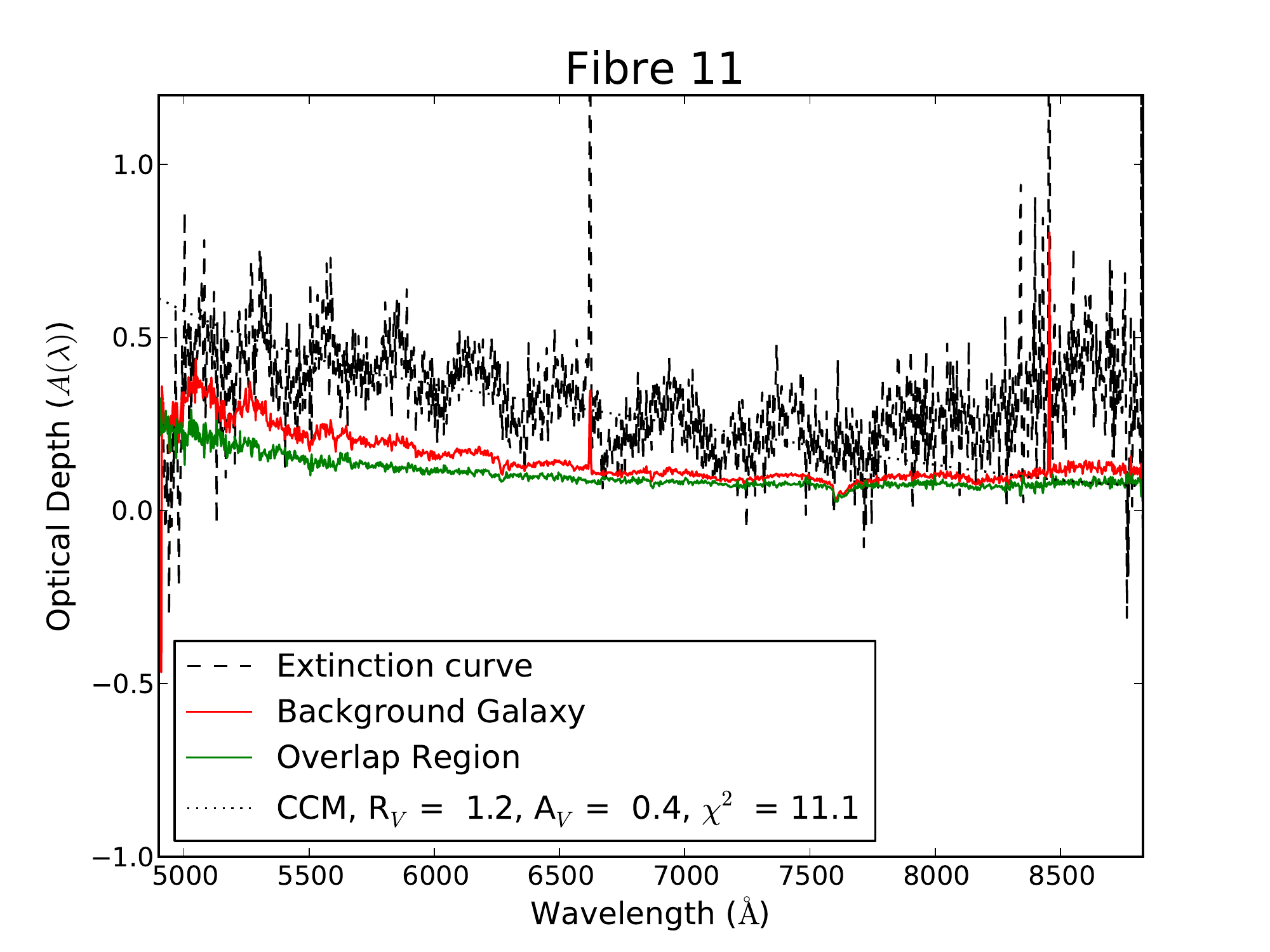}
\includegraphics[width=0.32\textwidth]{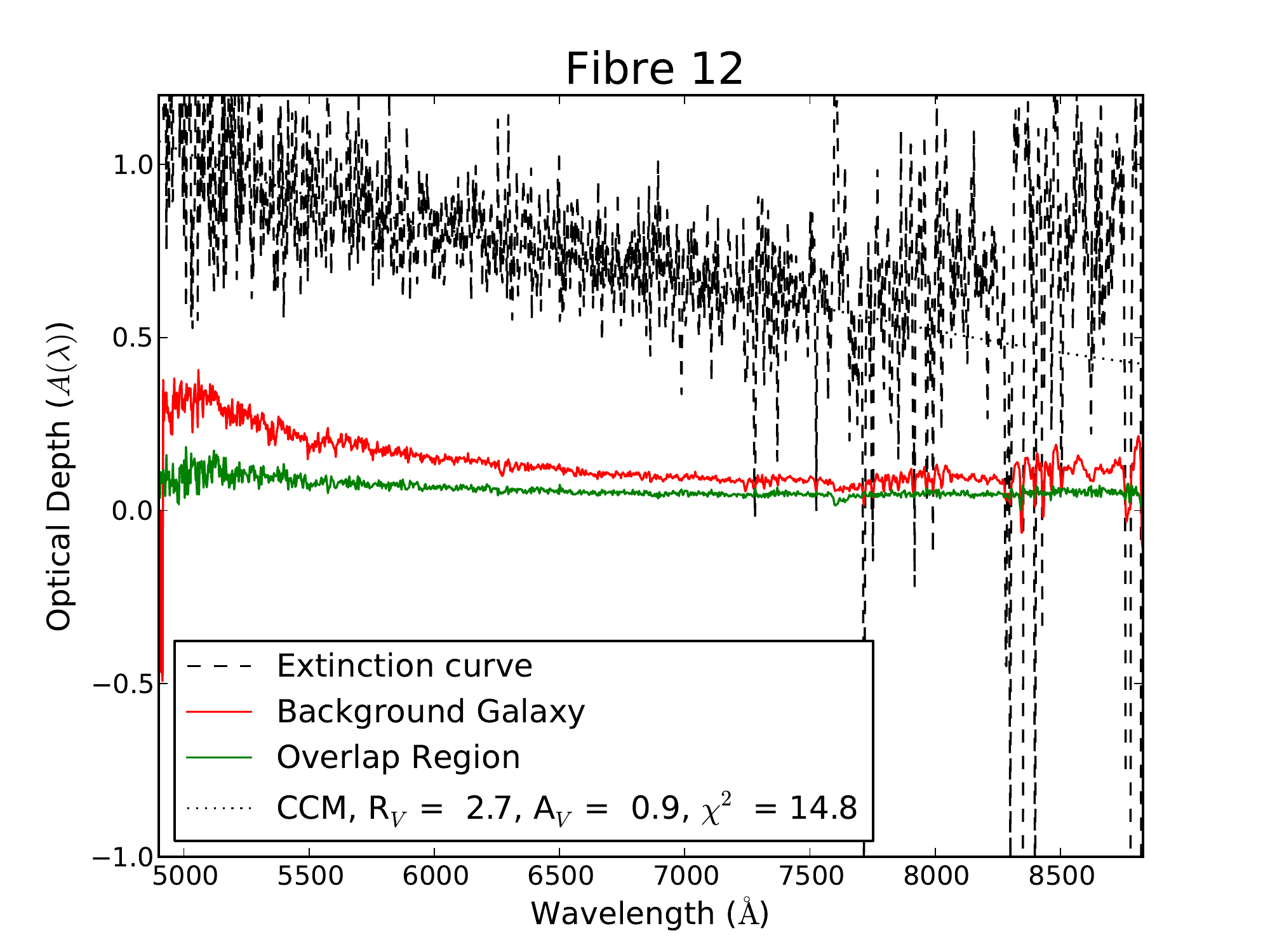}
\includegraphics[width=0.32\textwidth]{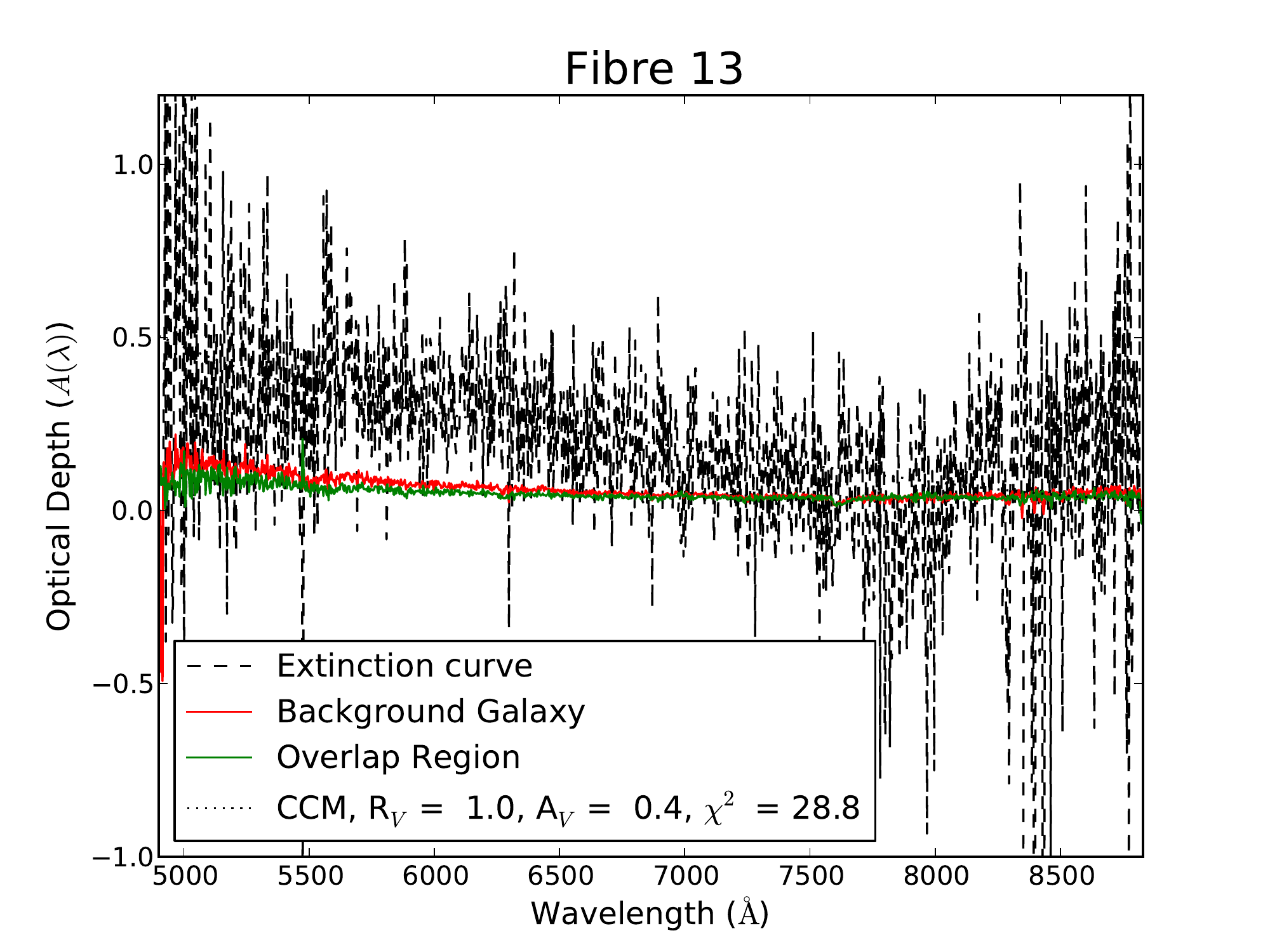}
\includegraphics[width=0.32\textwidth]{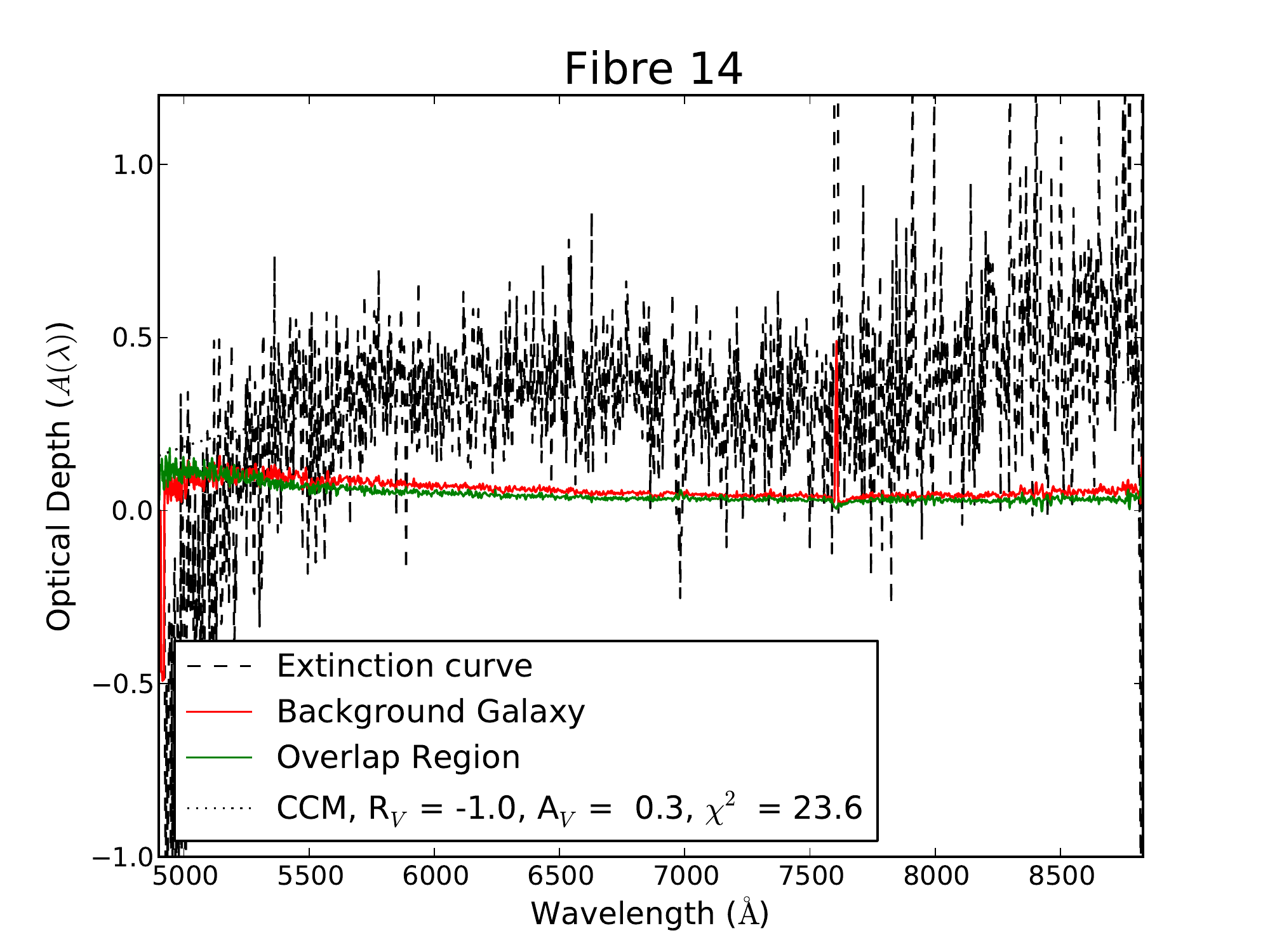}
\includegraphics[width=0.32\textwidth]{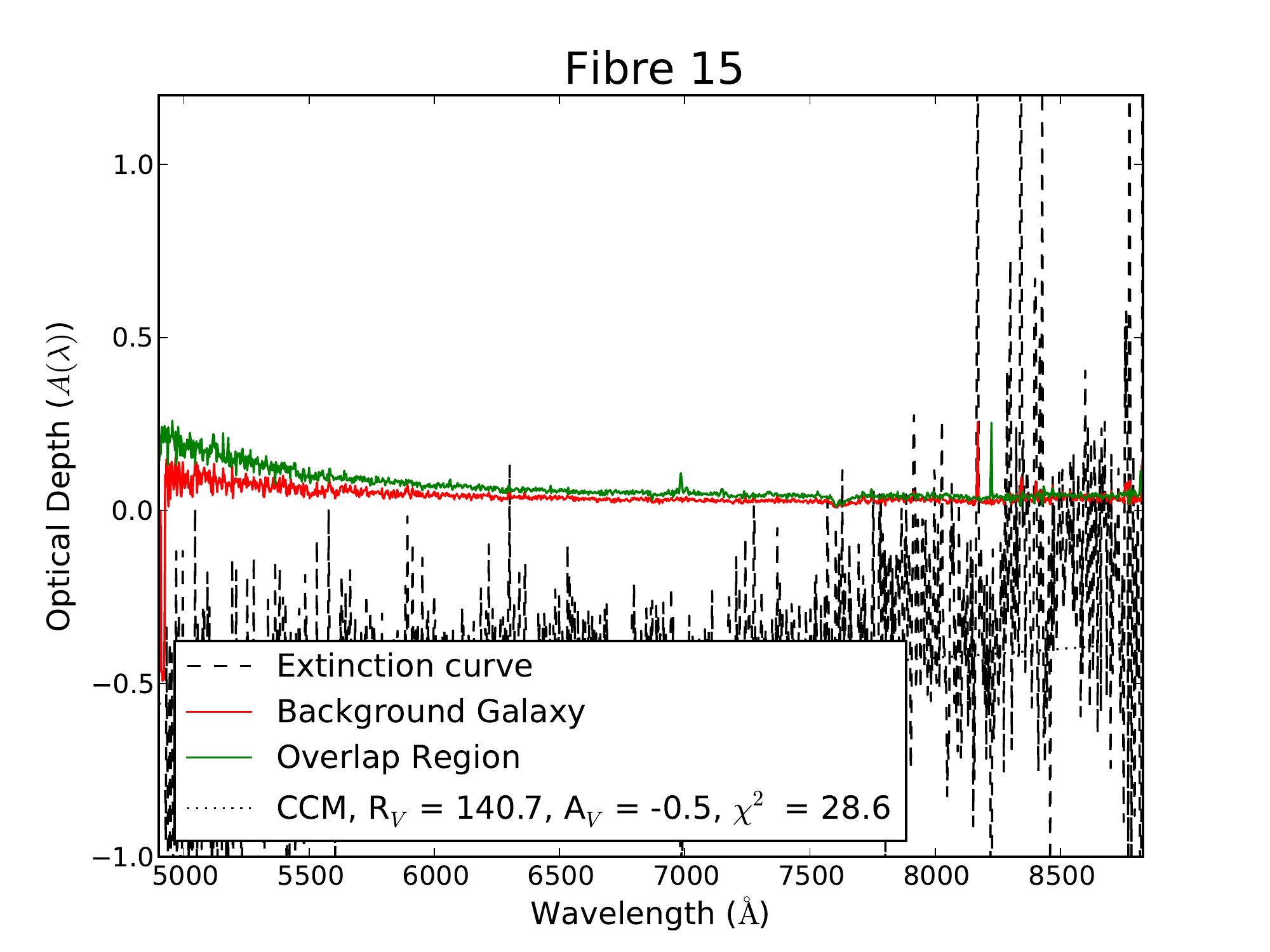}
\caption{The raw attenuation curves for Aperture II. The background (red) and the overlap (solid black line) spectra. The observed attenuation curve (dashed line) and the fit to this line (dotted line). }
\label{f:aper2:fib}
\end{center}
\end{figure*}
\end{document}